\documentclass[10pt]{iopart}
\usepackage{graphics, graphicx}%, amsmath, amssymb}
\usepackage{subfigure}
\usepackage[utf8]{inputenc}
\usepackage[usenames,dvipsnames]{color}

\usepackage{iopams}

\expandafter\let\csname equation*\endcsname\relax
\expandafter\let\csname endequation*\endcsname\relax
\usepackage{amsmath}
\usepackage{cite}

\definecolor{Red}{rgb}{1,0,0}

\newcommand{\ket}[1]{\ensuremath{\vert #1 \rangle}}

{\catcode`\|=\active
 \gdef\Braket#1{\begingroup
\mathcode`\|32768\let|\BraVert\left<{#1}\right>\endgroup}}
\def\BraVert{\egroup\,\mid\,\bgroup}

\newcommand{\re}{\mathrm{Re\,}}
\newcommand{\RR}{\mathbb{R}}

\begin{document}

\review{One decade of quantum optimal control in the chopped random basis}
\date{\today}

\author{Matthias M. M\"uller$^1$, Ressa S. Said$^2$, Fedor Jelezko$^2$, Tommaso Calarco$^{1,3}$,  and Simone Montangero$^{4,5}$}

\address{$^1$ Peter Grünberg Institute -- Quantum Control (PGI-8), Forschungszentrum J\"ulich GmbH, D-52425 Germany}

\address{$^2$ Institute for Quantum Optics \& Center for Integrated Quantum Science and Technology, Universit\"at Ulm, D-89081 Germany}

\address{$^3$ Institute for Theoretical Physics, University of Cologne, D-50937 Germany}

\address{$^4$ Dipartimento di Fisica e Astronomia ``G. Galilei", Universit\`a degli Studi di Padova \& INFN, Sezione di Padova, I-35131 Italy}

\address{$^5$ Padua Quantum Technology Center, Universit\`a degli Studi di Padova, I-35131 Italy}
\ead{ma.mueller@fz-juelich.de}

\begin{abstract}
The Chopped RAndom Basis (CRAB) ansatz for quantum optimal control has been proven to be a versatile tool to enable quantum technology applications, quantum computing, quantum simulation, quantum sensing, and quantum communication. Its capability to encompass experimental constraints -- while maintaining an access to the usually trap-free control landscape -- and to switch from open-loop to closed-loop optimization (including with remote access -- or RedCRAB) is contributing to the development of quantum technology on many different physical platforms. In this review article we present the development, the theoretical basis and the toolbox for this optimization algorithm, as well as an overview of the broad range of different theoretical and experimental applications that exploit this powerful technique.
\end{abstract}

\maketitle
%**************************************************************

%\tableofcontents

\section{Introduction}
A decade after its conception, the chopped random basis (CRAB) algorithm~\cite{Doria2011,Caneva2011,Rach2015} for optimal control of quantum systems (or Quantum Optimal Control -- QOC) has brought significant impact on many research areas in modern physics, especially the fields of quantum information, quantum technology and quantum many-body physics~\cite{Jurdjevic1996,DAlessandro2007,Brif2010,Glaser2015,Koch2016,Boscain2020}. In quantum information science and technology, QOC and the CRAB family of QOC algorithms in general, have demonstrated their capabilities on various physical platforms -- including nitrogen-vacancy centres in diamond~\cite{Scheuer2014,Waldherr2014,Dolde2014,Unden2016,Binder2017,Frank2017,Schmitt2017,Poggiali2018,Mueller2018}, trapped ions~\cite{Mueller2015,Fuerst2014,Pichler2016, Monz2011, Walther2012, Casanova2011, Singer2010, Zhang2018, Leibfried2003}, cold atoms~\cite{Rosi2013,Frank2014,Frank2016,Brouzos2015,Soerensen2018,Heck2018,Omran2019}, and superconducting qubits~\cite{Watts2015,Goerz2015,Hoeb2017} -- for numerous quantum optimization and information processing tasks, reaching from high performance sensing~\cite{Degen2017,Frank2017,Mueller2018,Poggiali2018,Rembold2020} to quantum metrology~\cite{Paris2004,Pezze2018,Mueller2015,Lovecchio2016,Pichler2016}, single-photon generation for quantum communication~\cite{Mueller2013,Duan2001,Ripka2018}, and state preparation and controlled unitary operations for  quantum computing and quantum simulation~\cite{Nielsen2010,Caneva2011b,Caneva2012,Goerz2015,Frank2014,Frank2016,Frank2017,Cui2017,Mueller2016}.
In quantum many-body physics~\cite{White1992,Schollwoeck2005,Schollwoeck2011,Brouzos2015,Alon2016,DeChiara2008,Doria2011,Caneva2011,Caneva2011b,Caneva2012,Caneva2014,Mueller2013, Frank2014, Mueller2016,Cui2017,Silvi2019,Weimer2019,Omran2019} QOC can shed light upon understanding and manipulating the dynamics of quantum many-body systems in a controlled way, including quantum phase transitions, an essential step toward advanced material sciences and the engineering of quantum technology platforms.
The key features of the CRAB ansatz for quantum optimal control are its capability to work under experimental constraints -- while at the same time being able to maintain favourable convergence properties by an access to the usually trap-free control landscape (see Sec.~\ref{sec:algorithm} and Refs.~\cite{Rabitz2006,Brif2010,Wu2012,Riviello2014,Rabitz2004,Hsieh2008,Ilin2018,Rach2015}) -- and its flexibility to switch from open-loop to closed-loop optimization including remote optimization~\cite{Rosi2013,Frank2017,Heck2018}.

\subsection{Structure of the Review}
In this review we present the development, the theoretical basis and the toolbox for this optimization algorithm, as well as an overview of a broad range of different theoretical and experimental applications exploiting this powerful technique.
The review is thus structured as follows. We first give an overview on the algorithm itself in Sec.~\ref{sec:algorithm}, including its structure, how and why it works, as well as software packages that implement or include it. In Sec.~\ref{sec:applications-toolbox}, we then present a toolbox that allows to tackle different classes of control problems like state preparation, entanglement generation or enhanced quantum measurements and other exemplary applications. In Sec.~\ref{sec:applications-per-platform}, we focus on the different physical systems and theoretical models that were investigated with CRAB by presenting different quantum technology platforms and QOC tasks that have been achieved on these platforms. To conclude, in Sec.~\ref{sec:conclusions} we give a summary and outlook.

\subsection{Quantum Optimal Control}
QOC has undergone a tremendous evolution over the last two decades helping to move quantum physics from the stage of observing and describing to the stage of engineering~\cite{Jurdjevic1996,DAlessandro2007,Brif2010,Glaser2015,Koch2016}. Originally, the field evolved in parallel to the nascent laser technology in the 1970s and 1980s in an attempt to control chemical reactions~\cite{Letokhov1977,Bloembergen1978,Brumer1986}. The monochromatic light of the laser allowed to slow down or enhance the chemical reactions by manipulating the energy levels or wavepackets of the reactants or by adding the energy needed for the reaction itself~\cite{George1982,Tannor1985,Kosloff1989}. Pulse shaping was introduced first by variational optimization~\cite{Tannor1985} and later by gradient algorithms~\cite{Kosloff1989}. A similar development was put forward in the field of nuclear magnetic resonance (NMR) by Rabitz and co-workers~\cite{Pierce1988,Shi1988}.
The currently widely used gradient-based QOC algorithms go back to an approach by Konnov and Krotov \cite{Konnov1999} which was introduced to the QOC community for instance by Sklarz et al. \cite{Sklarz2002}, and by Palao and Kosloff as Krotov algorithm~\cite{Palao2003} or in a variation as gradient ascent pulse engineering (GRAPE) algorithm by Khaneja et al.~\cite{Khaneja2005}. Recently, also QOC for open quantum systems has attracted increasing interest~\cite{Schmidt2011,Mukherjee2013,Hoyer2014,Kallush2014,Pawela2015,Mukherjee2015,Reich2015,Lovecchio2016,Koch2016}. QOC -- especially for open quantum systems -- has some overlap with the method of dynamical decoupling of the system environment modes, where a control pulse allows to enhance or suppress certain (un)desired interaction modes~\cite{Viola1998,Viola1999,Kofman2000,Kofman2001,Gordon2007,Biercuk2011,Green2013} and can also be engineered by optimization methods~\cite{Gordon2008,Clausen2012,Zwick2014,Mueller2018,Poggiali2018,Mueller2020}.

To successfully find a QOC solution, several conditions have to be fulfilled. A quantum system is \emph{controllable} if for every unitary transformation on the system state there is a control pulse that generates it~\cite{DAlessandro2007}. Nevertheless, this is not a necessary condition to find a QOC solution with high precision for a given transformation. Also, the transformations cannot be achieved arbitrarily fast for a finite energy of the control pulse (see section~\ref{sec:algorithm} for more details). This time-energy bound is known as the quantum speed limit (QSL)~\cite{Bhattacharyya1983,Margolus1998,Deffner2017} and its effect can be observed by QOC when the achievable control error sharply rises as the pulse operation time approaches the theoretical QSL~\cite{Caneva2009}. Similarly, the information contained in the control pulse has to be sufficient to steer the system to the desired target. Numerically, a so-called ``$2N-2$''-rule suggests that the number of degrees of freedom in the control pulse should at least correspond to the dimension $N$ of the quantum system~\cite{Moore2012,Caneva2014}. By quantifying the information content of the control pulse using Shannon's information theory~\cite{Shannon1948,Shannon1949} one can derive a relation between the bandwidth of the control pulse, the pulse operation time and the control error~\cite{Lloyd2014,Mueller2020,Gherardini2020}.

Furthermore, the experimental need for smooth pulses (e.g. bandwidth limited) gave rise to more sophisticated QOC algorithms that allowed to incorporate this constraint in different ways. Additional input came from the development of quantum many-body platforms where the numerical effort to simulate the system allowed to consider only a few parameters. This brought about the idea to make a bandwidth-limited few-parameter ansatz for the solution and CRAB was first introduced to optimize the preparation of a Mott insulator~\cite{Doria2011,Caneva2011}. For gradient-based QOC algorithms spectrally-limited versions of Krotov~\cite{Reich2014} and GRAPE~\cite{Motzoi2011} were obtained by incorporating filters into the gradient update. Alternatively, the CRAB ansatz can be used together with gradient-based QOC algorithms leading, e.g., to the GOAT algorithm (gradient optimization of analytic controls)~\cite{Machnes2018}. These gradient-based algorithms of the CRAB family are also called gradient optimization using parametrization (GROUP)~\cite{Soerensen2018}.

\subsection{Control Problem}\label{sec:control-problem}
\begin{figure}
	\centering
	\includegraphics[width=\textwidth]{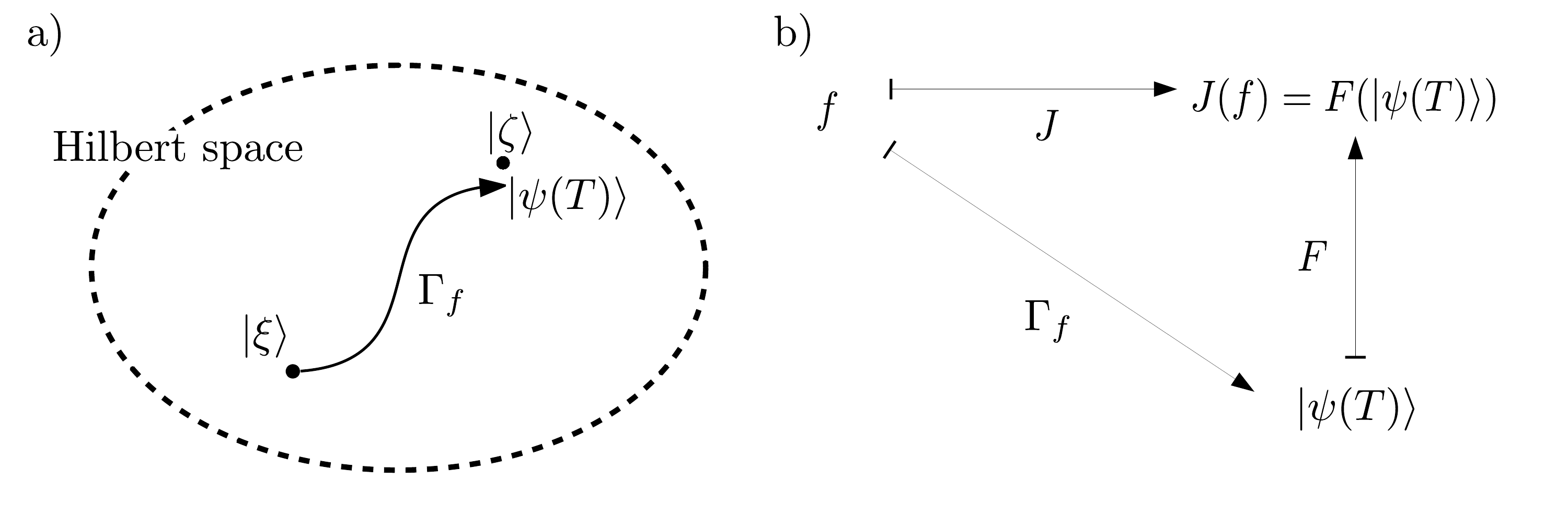}
	\caption{Schematic view of the control problem for a state to state transfer. The control pulse $f$ generates a time evolution $\Gamma_f$ that maps the initial state $|\xi\rangle$ onto a final state $|\psi(T)\rangle$ (ideally the target state $|\zeta\rangle$), panel a). The figure of merit or fidelity function $F$ evaluates the quality of the control (e.g., the overlap between final state and target state) and defines the control objective. Panel b) shows how a concatenation of time evolution and fidelity function allows to write the control objective directly as a function $J(f)$ of the control pulse $f$.}\label{fig:ControlProblem}
\end{figure}
To formalize a control problem in general terms we consider a system that evolves under controlled dynamics in the following sense. The dynamical equation of the system depends on a time dependent control pulse $f(t)$ (possibly also more than one) that determines the system evolution according to a map $\Gamma_f$ (see Fig.~\ref{fig:ControlProblem}). For example, we can consider an initial state $|\xi\rangle$ evolving as $|\psi(t)\rangle=\Gamma_f(t)(|\xi\rangle)=U_f(t)|\xi\rangle$ where the unitary $U_f(t)$ is the solution of the Schrödinger equation with Hamiltonian $H(t)= H_0 + f(t)H_1$, $H_0$ is the drift Hamiltonian and $H_1$ is the control Hamiltonian. In the example, we want to transfer the initial state to a target state $|\zeta\rangle$ at final time $T$ and thus set the control objective to be the overlap of the time evolved state and the target state: $F(|\psi(T)\rangle)=|\langle\zeta|\psi(T)\rangle|^2$. Here, $F(|\psi(T)\rangle)$ is the fidelity of the state transfer.
The control objective $J(f)$ can then be written implicitly as a function of only the control pulse $f(t)$ since the state of the system 
itself is a function of the control pulse, in the example $J(f)=F(|\psi(T)\rangle)$.
The control objective as a function of the control pulse is also called control landscape \cite{Brif2010,Wu2012,Moore2011,Riviello2014}.
The control problem is then to find a control pulse $f(t)$ that maximizes $J(f)$.
Additionally, $J(f)$ could also contain a term that depends only on $f(t)$, such as the pulse power $\frac{1}{T}\int_0^T f(t)^2 dt$. In section~\ref{sec:algorithm} we show how to tackle the control problem with the dCRAB algorithm and in section~\ref{sec:applications-toolbox} we introduce a broad variety of different control objectives.

%**************************************************************
\section{Algorithm}\label{sec:algorithm}
Originally, the CRAB algorithm was developed for quantum many-body systems whose time evolution can be efficiently simulated by time-dependent density matrix renormalization group (tDMRG)~\cite{White1992,Schollwoeck2005,Schollwoeck2011,DeChiara2008}. These systems were believed to be hardly tractable for control optimization using gradient-based algorithms \cite{Brif2010}, although recently a more efficient way for gradient-based optimization of many-body systems was introduced~\cite{Jensen2020}.
Nonetheless, QOC of many-body systems is attainable with a few parameter search approach~\cite{Doria2011,Caneva2011}. The key feature of the algorithm is to make an ansatz for the control function in a truncated basis (typically, but not exclusively, in the Fourier space) and the randomization of this truncated basis (e.g. randomization of frequency and phase of the Fourier basis functions).
By exploring in parallel several random instances of basis functions, with a relatively low number of parameters (basis functions) in a single optimization, a large portion of the original (unconstrained) function space can be explored.
Later, it was shown that by an iterative version (called dressed CRAB -- or dCRAB \cite{Rach2015}) of this optimization in a random basis the advantageous convergence properties of the unconstrained landscape \cite{Rabitz2004,Rabitz2006,Brif2010,Wu2012,Riviello2014,Ilin2018} can be maintained also in a constrained (e.g. bandwidth-limited or intensity-limited) setting.
A further development of the algorithm is the closed-loop application of CRAB on an experimental set-up~\cite{Rosi2013,Frank2017} including with remote access over the internet, the so-called RedCRAB algorithm \cite{Heck2018}.
In this section we explain in detail the different developments of the CRAB algorithm, the possible choices of basis, underlying optimization methods
and the relation between convergence, efficient simulation and required resources~\cite{Lloyd2014,Mueller2020}.

\subsection{Control Problem revisited}\label{sec:control-problem-simple}
Hereafter, we consider the optimization of a state transfer problem where the figure of merit is the overlap of the final state after the time evolution with the target state. Other control problems and their mathematical formulation will be discussed in Section~\ref{sec:applications-toolbox}.

In this exemplary case the control objective reads
\begin{eqnarray}\label{eq:fidelity-statetostate}
	J(f)=F(|\psi(T)\rangle)=|\langle\zeta|\psi(T)\rangle|^2\,,
\end{eqnarray}
where $|\zeta\rangle$ is the target state and the time evolution is given by the Schr\"odinger equation
\begin{eqnarray}
	i\frac{\partial}{\partial t}|\psi(t)\rangle=\left(H_0+f(t)H_1\right)|\psi(t)\rangle\nonumber\\
	|\psi(0)\rangle=|\xi\rangle\,.
\end{eqnarray}
The control pulse $f(t)$ is determined by the optimization algorithm in order to maximize the control objective. If no particular constraints are considered, $f(t)$ is taken, e.g., from the function space $L_2$ of square-integrable functions or from the space of measureable essentially bounded functions $L_\infty$ (see also \cite{Boscain2020}) -- that is, potentially an infinite dimensional space has be to searched.

\subsection{Key Idea of the CRAB Algorithm}
\begin{figure}
	\centering
	\includegraphics[width=0.5\textwidth]{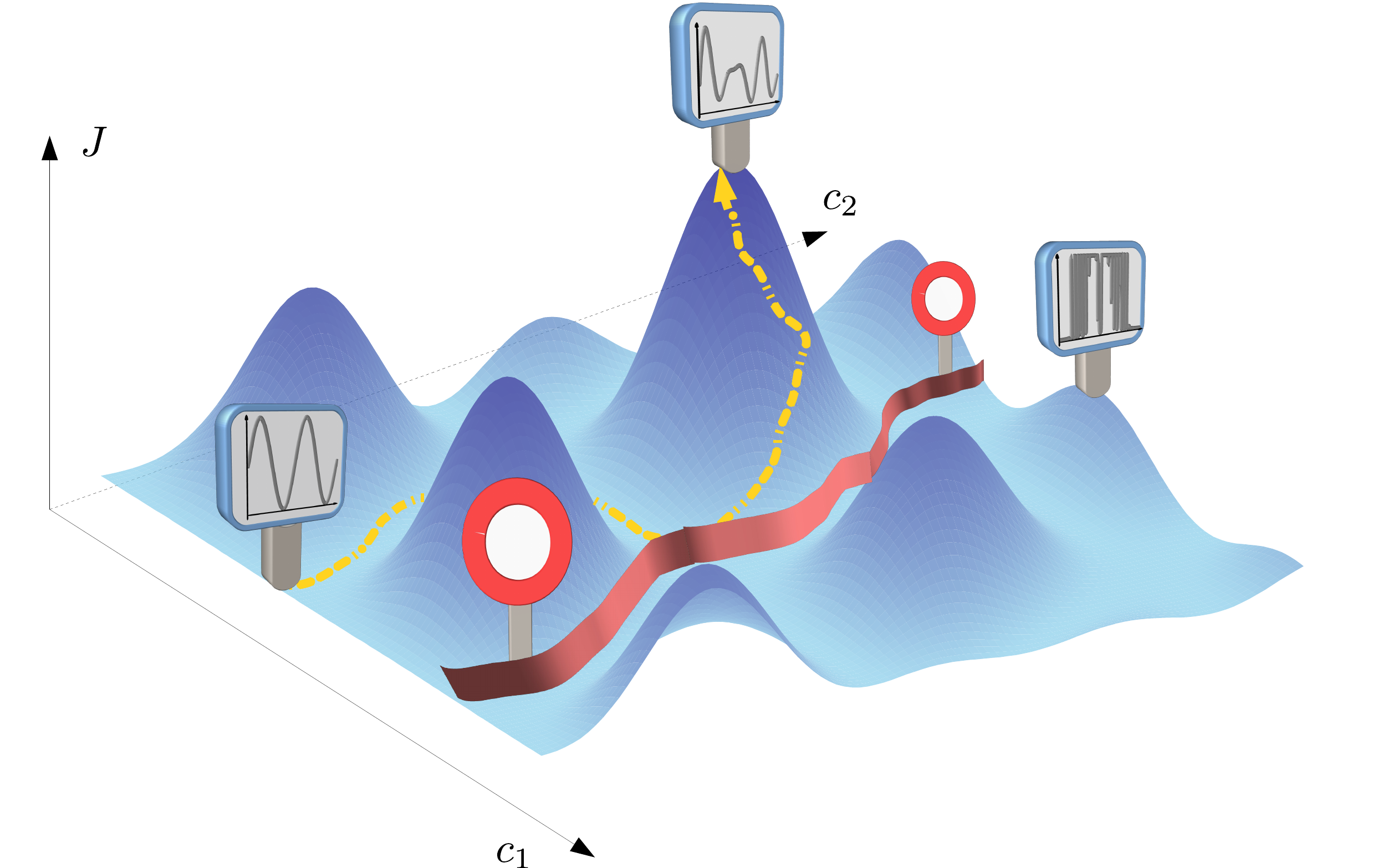}%
	\includegraphics[width=0.5\textwidth]{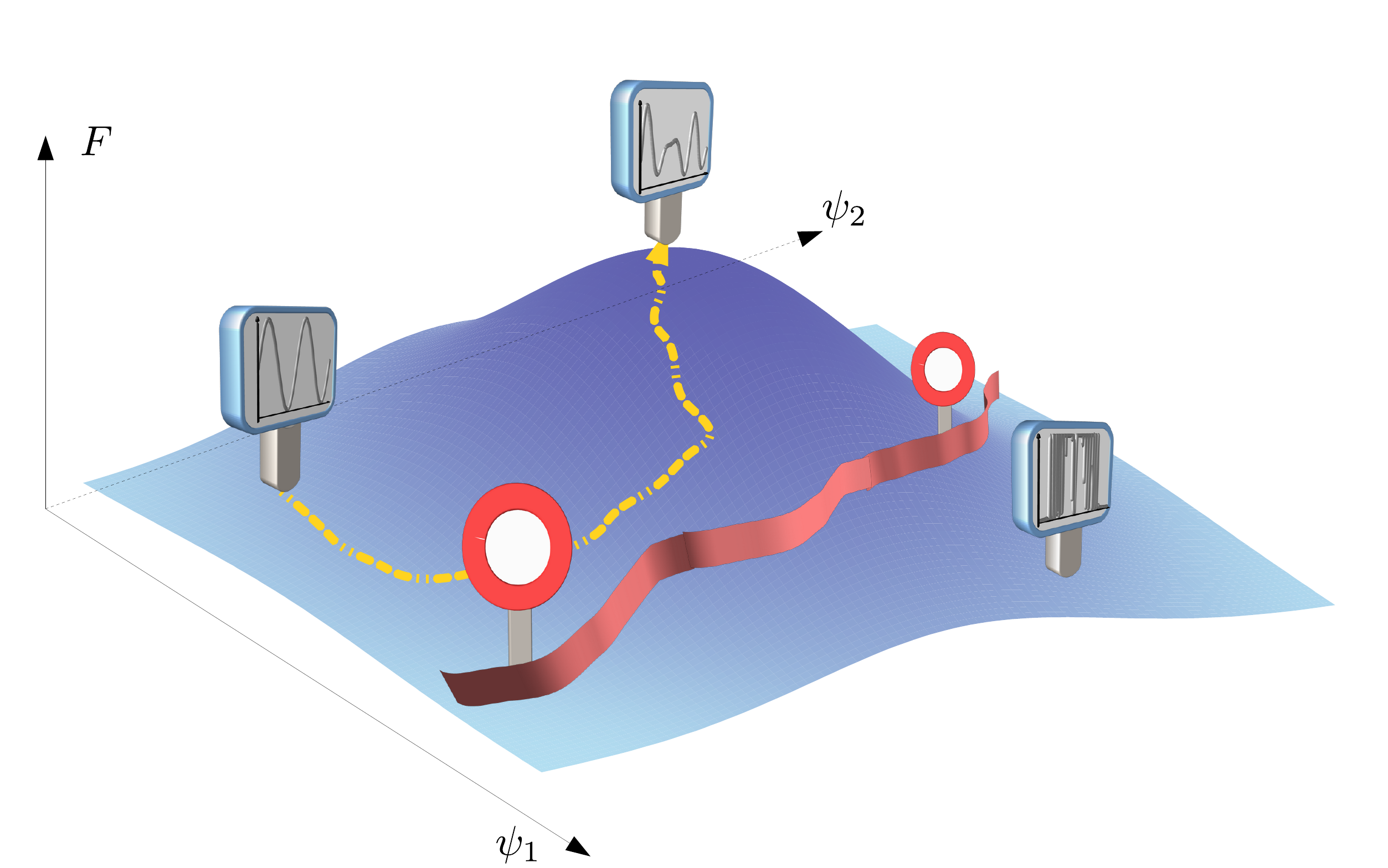}
	\caption{Schematic view on the control landscape and optimization of a control problem. Left panel: By varying the control parameters $c_1$ and $c_2$ one can change the control pulse $f$ and maximize the control objective $J(f)$. The control landscape shows local maxima and some areas might be prohibited due to constraints on the control pulse (the red barrier can not be crossed). The functional $J(f)$ is also called dynamic control landscape. Right panel: Here, the fidelity $F$ is plotted as a function of the final time state (in the figure given by its two coordinates $\psi_1$ and $\psi_2$). It usually is a simple function with one maximum. The function $F(|\psi(T)\rangle)$ is also called kinematic control landscape.}\label{fig:landscape}
\end{figure}

The key idea of the CRAB algorithm to maximize $J(f)$ is to expand the control pulse $f(t)$ in a truncated basis with $N_c$ basis functions $f_{i}(t)$ ($i=1,\dots\, N_c)$:
\begin{eqnarray}\label{eq:CRAB1}
	f(t)=\sum_{i=1}^{N_c}c_{i}f_{i}(t)\,.
\end{eqnarray}
The optimization is then performed on this subspace of reduced dimension, or in other words, an optimal set of coefficients $c_{i}$ ($i=1,\dots\, N_c)$ has to be found. This can be done by standard direct search algorithms -- e.g. the Nelder-Mead simplex algorithm~\cite{Nelder1965}, Powell's method \cite{Powell1964}, subspace simplex search \cite{Rowan1990}, parallel simplex~\cite{Lee2007}, particle swarm optimization or hybrid particle swarm and simplex methods \cite{Fan2004}, Monte Carlo optimization (Metropolis or simulated annealing~\cite{Metropolis1953}) or covariance matrix adaption evolution strategy (CMA-ES)~\cite{Hansen2003} -- or by calculating the gradient with respect to the parametrization leading to algorithms like GOAT~\cite{Machnes2018} or GROUP~\cite{Soerensen2018}. The left panel of Figure~\ref{fig:landscape} shows an example of such an optimization for two parameters. A variation of these two parameters will cause a variation of the control pulse $f(t)$ and thus of the control objective. The control problem is reduced to a search for the maximum of the control objective $J(f)=J(c_1,c_2)$, i.e., a simple search in two dimensions.

If one wants to start optimization from a guess pulse $g(t)$ that cannot be exactly expanded in the basis, this can be done by setting
\begin{eqnarray}\label{eq:CRAB-guess}
	f(t)=g(t)\Big(1+\sum_{i=1}^{N_c}c_{i}f_{i}(t)\Big)\,.
\end{eqnarray}
However, the restriction of the search basis to $N_c$ dimensions given by the CRAB expansion in Eqs.~\eqref{eq:CRAB1} and \eqref{eq:CRAB-guess}, introduces a constraint on the search space that might lead to a non-optimal solution, i.e. the algorithm is trapped in a local minimum arising due to the constraint -- a so-called false trap~\cite{Moore2011}. A first step to weaken the constraint is the randomization of the basis functions: a standard choice for the basis functions are trigonometric functions, often multiplied by a shape function $1/\Lambda(t)$ that fixes the pulse to a given value at initial and final time:
\begin{eqnarray}\label{eq:CRAB2}
	f(t)=g(t)\Big(1+\sum_{i=1}^{N_c/2}c_{i}\frac{\cos(\omega_{i}t)}{\Lambda(t)}+\sum_{i=N_c/2+1}^{N_c}c_{i}\frac{\sin(\omega_{i}t)}{\Lambda(t)}\Big)\,.
\end{eqnarray}
The frequencies $\omega_{i}$ are then chosen randomly around the principal harmonics within some interval $[0,\omega_{\mathrm{max}}]$. Different choices of basis functions are possible, e.g., generalized Chebyshev polynomials where the (non-integer) order of the polynomial is chosen randomly from an interval. All admissible basis functions together span a space $\mathcal{C}\subset L_2$, the space of admissible controls. In principle, every subspace of $L_2$ that can be parameterized in an appropriate way can be used.
Now, for each set of random functions/frequencies a different $N_c$-dimensional subspace of $\mathcal{C}$ is spanned and therefore an optimization of the coefficients explores a different section of the potentially infinite dimensional space of admissible controls $\mathcal{C}$. This is the core strength of this algorithm and the power of the randomization was demonstrated numerically already when the algorithm was first introduced in Refs.\,\cite{Doria2011,Caneva2011}.

\subsubsection{From CRAB to dCRAB}\label{sec:DCRAB}
As we have seen, the choice of admissible basis functions limits the space of admissible controls to a subspace $\mathcal{C}\subset L_2$. A specific set of $N_c$ basis functions for the CRAB expansion, however, can span only the even smaller ($N_c$-dimensional) space $\mathrm{span}\{f_i\}\subset \mathcal{C}$. While the restriction to $\mathcal{C}$ can reflect a physical constraint (e.g., limited bandwidth of the controls), the further constraint given by the specific random instance of $N_c$ basis functions is merely technical and if possible should be circumvented to improve convergence of the algorithm. Indeed, such an additional constraint could introduce additional traps (false traps) in the control landscape. To overcome this problem, and escape from such false traps, one can introduce in an iterative way new random basis functions and use the optimal solution from the previous optimization as a guess pulse for the next iteration. From here on, we call these iterative basis changes super-iterations to distinguish them from the iterative optimization of the coefficients during one super-iteration. Then, in the $j$-th super-iteration 
one optimizes the coefficients $c_i^j$ ($i=0,\dots, N_c)$ of
\begin{eqnarray}\label{eq:dCRAB}
	f^j(t)=c_0^j f^{j-1}(t)+\sum_{i=1}^{N_c}c_i^j f_i^j(t)\,,
\end{eqnarray}
where $f_i^j(t)$ are the new basis functions, and $f^{j-1}(t)$ is the control pulse obtained from the $j-1$th iteration. The coefficient $c_0^j$ allows the optimization to move along the direction of the old pulse, while the coefficients $c_i^j$ ($i=1,\dots,N_c$) allow it to move along the new search directions $f_i^{j}(t)$.
As a consequence, in each super-iteration the old pulse is dressed with new search directions and this procedure was named dressed Chopped Random Basis (dCRAB) algorithm~\cite{Rach2015}.
This updating of the search directions can also be understood as an extension of Powell's method \cite{Powell1964} to an infinite dimensional search space: while in Powell's method in a finite dimensional search space after reaching a local optimum the search directions are updated through linear combinations of the original search directions, here the search directions are replaced by new ones out of the infinite dimensional search space.

In the following section we will shortly review the theory of control landscapes reviewing why this iterative update indeed allows the optimization to escape from potential false traps.

\subsubsection{Control Landscapes in the chopped random basis}\label{sec:landscapes}
The control landscape $J(f)=F(|\psi(T)\rangle)$ can be looked at as a function of the control pulse $f(t)$ or as a function of the final state $|\psi(T)\rangle$ \cite{Brif2010,Wu2012,Moore2011,Riviello2014}. Figure \ref{fig:landscape} gives a schematic view on how a control problem looks like in the dynamic and in the  kinematic formulation, that is in terms of the control objective as a function of the control pulse $J(f)$  (left panel) and as a function of the final state $F(|\psi(T)\rangle)$ (right panel), respectively. Optimization usually leads to so-called critical points of the landscape, that is, points fulfilling the condition
\begin{eqnarray}\label{eq:criticalpoint-chainrule}
	\delta J=\langle \nabla F(\psi(T))|\delta \psi(T)\rangle=0\quad\forall\;\;\delta f\,.
\end{eqnarray}
The critical points can be divided into singular and regular critical points, where regular
means that for every $|\delta\psi(T)\rangle$ in the Hilbert space of the problem there is a
change in the control $\delta f$ that generates it. Instead, for singular points this is not the case.
Furthermore, a critical point is either kinematic or non-kinematic, where kinematic means that vanishing $\delta J$ implies a vanishing gradient of the fidelity $\nabla F(\psi(t))=0$. One can see immediately that regular critical points are always kinematic since for regular critical points the gradient of the fidelity is orthogonal to the whole Hilbert space according to Eq.~\eqref{eq:criticalpoint-chainrule} and thus vanishes. Instead, singular critical points can be kinematic or non-kinematic.

Typically, $F(|\psi(T)\rangle)$ is a very simple function and thus a vanishing gradient $\nabla F(\psi(T))=0$ corresponds to the global maximum or a saddle point. 
For example, the state transfer with control objective Eq.~\eqref{eq:fidelity-statetostate} has a vanishing gradient only at the global maximum and the global minimum, while for the fidelity of a $SU(N)$ gate there are the global maximum, the global minimum and $N-1$ saddle points \cite{Riviello2014}. Thus, kinematic critical points, i.e., the ones where $\delta J=0\,\forall\,\delta f$ coincides with a vanishing gradient $\nabla F(\psi(T))=0$, are not traps and we have to be concerned only about the non-kinematic critical points as potential traps. However, at least for controllable systems, there seem to be mainly regular (and thus kinematic) critical points and the singular (and potentially non-kinematic) critical points typically are no big threat \cite{Wu2012,Riviello2014,Rabitz2004}. A mathematically rigorous proof for this statement, however, so far has been achieved only for a few simple examples like two-level systems or transmission through a barrier, see Ref.~\cite{Ilin2018}.

The above statements about regularity, however, assume that the control $f$ is a function of $L_2$, whereas for CRAB the optimization operates in the smaller space $\mathrm{span}\{f_i\}\subset \mathcal{C}\subset L_2$. This means that potentially one encounters pseudo-critical points with
\begin{eqnarray}\label{eq:pseudocriticalpoint}
	\delta J=0\;\forall\;\delta f=\sum_{i=1}^{N_c}f_i(t)\delta c_i\quad\text{but}\nonumber\\
	\exists\;\delta f\;:\;f+\delta f\in\;L_2\;,\;\delta J\neq 0\,.
\end{eqnarray}
These points are called false traps~\cite{Moore2011} as they arise only artificially from the choice of the basis and their influence can hinder the convergence of the algorithm~\cite{Moore2011,Riviello2015}.  In the following we will show how the super-iterations of dCRAB, Eq.~\eqref{eq:dCRAB}, can open a way out of these false traps:

\paragraph{Removing false traps for CRAB}
From Eqs.~\eqref{eq:criticalpoint-chainrule} and \eqref{eq:pseudocriticalpoint} follows that, for a pseudo-critical point the gradient is such that $\langle \nabla F(\psi(T)) | \cdot\rangle=\mathrm{Re}\langle \phi_T | \cdot\rangle$ for some non-zero vector $\phi_T$.
In the typical case of
\begin{eqnarray}
	i\frac{\partial}{\partial t}|\psi(t)\rangle=\left(H_0+f(t)H_1\right)|\psi(t)\rangle\,,\nonumber\\
	|\psi(0)\rangle=|\xi\rangle\,,\nonumber\\
\end{eqnarray}
a perturbative analysis yields
\begin{eqnarray}
	|\delta\psi(T)\rangle=-i U(T)\int_0^T U^\dagger (t)H_1U(t)|\xi\rangle\delta f(t)\mathrm{d}t\\
	\mathrm{Re}\langle\phi_T|\delta\psi(T) \rangle
	=\int_0^T k(t)\delta f(t)\mathrm{d}t \\
	k(t)=-\mathrm{Im}\langle \phi_T|U(T) U^\dagger (t)H_1U(t)|\xi\rangle
\end{eqnarray}
where $k$ is a continuous function and $k\neq0$ since $k=0$ violates Eq.~\eqref{eq:pseudocriticalpoint}.
Note that $|\phi(t)\rangle = U(t)U^{\dagger}(T)|\phi_T\rangle$ is also called the backward propagated or adjoint state and plays a central role in gradient-based optimization methods~\cite{Konnov1999,Palao2003,Khaneja2005}.
If we choose a new direction $\delta f(t)=\sin(\omega_r t)\delta c$ with a new random frequency $\omega_r$  we get almost surely
\begin{eqnarray}\label{eq:almostsureoverlap}
	\int_0^T k(t)\delta f(t)\mathrm{d}t\neq 0\,,
\end{eqnarray}
where `almost surely' in probability means that in principle the integral can vanish (e.g., if accidentally the new frequency was already present in the expansion) but the probability for this to happen is 0. Perturbing the control pulse $f$ by this new frequency contribution we find
\begin{eqnarray}\label{eq:outofartificaltrap}
	\delta J =\langle \nabla F(\psi(T))|\delta \psi(T)\rangle\neq 0
\end{eqnarray}
and thus  the false trap has been removed. The recipe to escape from a false trap is thus to add a new random frequency term to the CRAB expansion once the optimization is stuck in a false trap (or in other terms a new randomly picked basis function out of the chosen set of admissible basis functions $\mathcal{C}$). This can be done also without increasing the total number of coefficients since once in the false trap for a given set of basis functions and coefficients there is no more use in varying the old coefficients, and they can be kept at their constant value. This insight leads to the iterative update of the basis functions (see Eq.~(\ref{eq:dCRAB}) and Ref.~\cite{Rach2015}).

As a side remark, note that if the function $k(t)$ is chosen as pulse update, it moves the state along the gradient of the (kinematic) control landscape. If this kinematic control landscape has dimension $2N-2$, at each iteration one can find $2N-2$ functions that form a basis for $k(t)$, the update in the gradient direction \cite{Hsieh2008}. This suggests that generally the number of degrees of freedom in the control pulse and the dimensionality of the kinematic control landscape (i.e. the control problem in the state space) have to be at least the same to solve the control problem. This fact has been confirmed numerically~\cite{Moore2012,Caneva2014}, and more details are provided in Sec.~\ref{sec:constrained-optimization}. For dCRAB, one can show that $2N-2$ random directions can generate a pulse update that locally moves the state into the direction of the gradient, thus reproducing the result of \cite{Hsieh2008} but with the additional constraint on the bandwidth~\cite{Rach2015}.

\subsection{Constrained Optimization}\label{sec:constrained-optimization}
As already mentioned, the CRAB expansion naturally introduces a constrained bandwidth. Other constraints can potentially be implemented by introducing a different basis for the expansions. A very common constraint for optimization (often together with the bandwidth constraint), however, is a constraint on the power or energy of the control pulse. A standard way to implement this constraint is through Lagrange multipliers or penalty terms. Such a penalty term in the control objective can constrain the pulse height
\begin{eqnarray}\label{eq:pulseheight-weight}
	J=F-\lambda \max_{t}|f(t)|\,,
\end{eqnarray}
or the pulse energy:
\begin{eqnarray}\label{eq:pulsepower-weight}
	J=F-\lambda \int_0^T f(t)^2 dt\,.
\end{eqnarray}
Another possibility is to limit the pulse height to a maximum value $f_{\mathrm{max}}$ by a hard wall constraint. This can be achieved by using the update formula
\begin{eqnarray}\label{eq:pulseheight-wall}
	\tilde{f}^j(t)=c_0^jf^{j-1}(t)+\sum_{i=1}^{N_c}c_i^j f_i^j(t)\,,\\
	f^j(t)=\begin{cases}
		\tilde{f}^{j}(t)&\text{if}\,\,|\tilde{f}^{j}(t)|<f_{\mathrm{max}}\\
		\mathrm{sgn}(f^j(t)) f_{\mathrm{max}}&\text{otherwise}.
	\end{cases}
\end{eqnarray}
As opposed to the penalty terms, the hard wall constraint potentially increases the bandwidth of the control pulse. Numerical experiments \cite{Rach2015} indicated that the maximal height of the control pulse could be decreased by about one order of magnitude compared to the unconstrained case without interfering with the success of the optimization. In this scenario, the hard wall approach could decrease the pulse height by about an additional factor of 3 compared to Eq.~\eqref{eq:pulseheight-weight}.
Alternatively, if one desires to maintain the bandwidth constraint while introducing an additional hard wall constraint, one can achieve this by rescaling the control pulse at each iteration of the search algorithm according to the update formula
\begin{eqnarray}\label{eq:pulseheight-wall2}
	\tilde{f}^j(t)=c_0^jf^{j-1}(t)+\sum_{i=1}^{N_c}c_i^j f_i^j(t)\,,\\
	\tilde{f}^j_{\mathrm{max}}=\max_{t} |\tilde{f}^j(t)|\,,\\
	f^j(t)=\begin{cases}
		\tilde{f}^{j}(t)&\text{if}\,\,\tilde{f}^j_{\mathrm{max}}<f_{\mathrm{max}}\\
		\frac{f_{\mathrm{max}}}{\tilde{f}^j_{\mathrm{max}}}\tilde{f}^j(t) &\text{otherwise}.
	\end{cases}
\end{eqnarray}
In this way, the pulse maximum (absolute) value is exactly $f_{\mathrm{max}}$ but its shape is not affected.

For constrained optimization it is especially important to consider the resources required to solve a QOC problem. Whether or not a solution can be found, not only depends on the controllability of the system~\cite{DAlessandro2007,Turinici2003}, but also on the given resources:  the finite pulse operation time, the finite pulse power and finally the degrees of freedom of the control pulse (e.g., given by the bandwidth or parametrization of the control pulse).
The necessary condition on the first two resources can be formulated as a time-energy bound similar to the Heisenberg uncertainty relation and is commonly known as the Quantum Speed Limit (QSL)~\cite{Deffner2017,Bhattacharyya1983,Margolus1998,Caneva2009,Deffner2013}.
For a state transfer in a two-level system with constant Hamiltonian and initial energy variance $\Delta E$, the QSL assumes the form of the Bhattacharyya bound~\cite{Bhattacharyya1983} which was also observed numerically with QOC  solutions~\cite{Caneva2009}:
\begin{equation}
	T_\text{QSL}\ge \Delta E^{-1} \arccos|\langle\zeta|\xi\rangle|\,.
\end{equation}
This bound can be generalized to more complex system dynamics, including for open systems, where essentially the energy variance has to be substituted by the Schatten p-norm of the dynamical operator describing the system~\cite{Deffner2013,Deffner2017}. In Sec.~\ref{sec:many-body} we will further discuss the QSL in terms of the fastest path connecting two states of a quantum many-body system.

\begin{figure}
	\centering
	\includegraphics[width=0.48\textwidth]{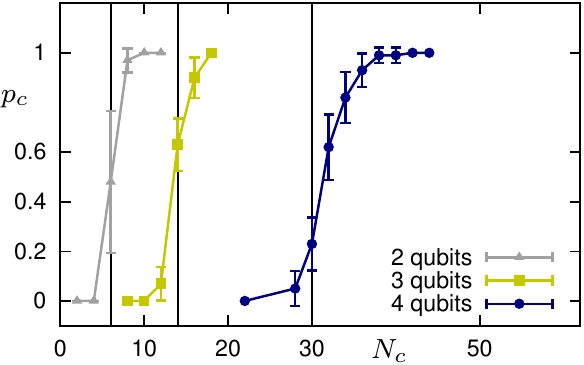}%
	\qquad%
	\includegraphics[width=0.48\textwidth]{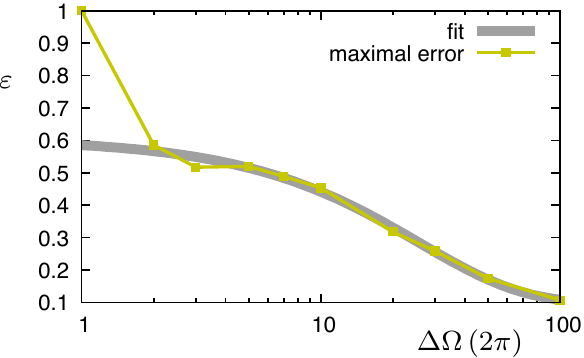}
	\caption{Constraints and Information Content. Limiting the degrees of freedom of the control pulse acts as a constraint on the optimization. In the noiseless case (left panel, taken from Ref.~\cite{Rach2015}), the degrees of freedom ($N_c$ in the case of CRAB) have to follow the "$2N-2$" rule (indicated by the vertical black lines for the three different system sizes) in order to obtain a high success probabilty $p_c$ of the optimization. Right panel: For dCRAB the degrees of freedom are not limited by the number of coefficients $N_c$ in a single super-iteration, but by the allowed bandwidth $\Delta\Omega$ of the control pulse. Information theory allows to link the control error to the capacity of a noisy channel and derive a scaling law (grey line) for the control error that can be observed also numerically (yellow squares), as in this case of a single qubit with a decaying level.}\label{fig:constraints}
\end{figure}
To quantify the influence of the restricted search space, we consider the dimension $D_{f}$ of the search space or space of admissible controls $\mathcal{C}$. It is given by the bandwidth limit of the CRAB expansion, or more generally by the number of independent degrees of freedom in the control pulse. From an information theoretical argument \cite{Lloyd2014} it was shown how the number of degrees of freedom $D_f$ needed to accomplish a control task depends on the dimension $D_s$  of the system space (or more precisely the dimension $D_r$ of the control problem) and on the tolerated error $\varepsilon$~\footnote{To understand better the role of $D_r$, we consider the example of a state transfer: in an $N$-dimensional complex Hilbert space the real dimension is $D_s=2N$. If we consider only physical states (with norm 1) and disregard the global phase, we obtain the dimension of the control problem $D_r=2N-2$. This coincides with the so-called ``$2N-2$"-rule of Refs.~\cite{Moore2012,Caneva2014}. More examples are discussed in Sec.~\ref{sec:applications-toolbox}.}.
In detail, using Shannon's theory of information~\cite{Cover2005,Shannon1948,Shannon1949} one can consider the control problem as a communication channel between the control pulse (input) and the final state of the system (output). The information that can be transmitted through such a channel, $I_{f}\equiv CT$, is the product of the channel capacity $C$ and the pulse operation time. For a noiseless channel with a finite bit depth $f_{\mathrm{max}}/\Delta f$ of the control pulse (where $f_{\mathrm{max}}$ is the maximally allowed modulation and $\Delta f$ the resolution of the control pulse) and a bandwidth $\Delta\Omega$ of the control pulse, the channel capacity is given by Hartley's law
\begin{equation}
	C = \Delta\Omega \log_2 \left(1 + \frac{f_{\mathrm{max}}}{\Delta f}\right).
\end{equation}
For a noisy channel, one has to consider the power $P_f=\frac{1}{T}\int_0^T f(t)^2 dt$ of the conrol pulse and its power spectral density $\hat f(\omega)$ as well as the power $P_n$ of the noise and the power spectral density of  the noise $\hat n(\omega)$. For Gaussian white noise the channel capacity is then
\begin{equation}\label{S-H-theorem}
	C= \Delta\Omega \log_2 \left(1 + \frac{P_f}{P_n}\right)\,,
\end{equation}
while in the more general case of colored noise it is given by
\begin{equation}\label{Shannon_law}
	C= \int_{0}^{\omega_{\rm max}} \log_2 \left(1 + \frac{\hat{f}(\omega)}{\hat{n}(\omega)}\right)d\omega\,.
\end{equation}
We now consider the information needed to steer the system, i.e., the amount of information that we have to transmit over the channel: it depends on the desired precision and reads $-D_r\log_2{\varepsilon}$. The information contained in the pulse has to be at least equal. As a consequence, a bound on the error of the control problem is obtained:
\begin{equation}\label{eq:bound_error}
	\varepsilon \geq 2^{-\frac{I_{f}}{D_r}}.
\end{equation}
This bound can also be cast into a time bound which specifies the time needed to transmit the necessary information to steer the system at given bandwidth:
\begin{equation}\label{eq:bound_T}
	T \geq -\frac{D_r}{C}\log_{2}(\varepsilon).
\end{equation}
These bounds were first found by Lloyd et al.~\cite{Lloyd2014} and were later generalized to colored noise~\cite{Mueller2020} and to the case where not only the system is quantum but also the controller~\cite{Gherardini2020}.

In the noiseless case, the error bound confirms the heuristic ``$2N-2$"-rule that was confirmed numerically in Refs.~\cite{Moore2012,Caneva2014,Frank2016,Rach2015,Mueller2016}. Fig.~\ref{fig:constraints} shows two examples for this information theoretical bound: The left panel is taken from Ref.~\cite{Rach2015} and shows the performance of CRAB in terms of the success probability $p_c$ of an optimization as a function of the number of  random basis functions $N_c$. In the reference a random Ising model with 2, 3 and 4 qubits was studied and the optimization of a state transfer was considered successful when it managed to overcome a certain threshold for the control error ($\varepsilon<10^{-3}$). The success probability was then calculated over 10 different random initial and target states and 100 trials per random instance. The vertical lines indicate the ``$2N-2$"-rule for the three different system sizes.
For dCRAB we do not have this dependence on $N_c$: due to the super-iterations, the relevant constraint on the pulse is its bandwidth (or the restriction to the space of admissible controls $\mathcal{C}$). The right panel of  Fig.~\ref{fig:constraints} shows the control error as a function of the bandwidth of the control pulse for a state transfer in a single qubit subjected to decoherence, in particular a decay channel. This result is taken from Ref.~\cite{Mueller2020}, where an optimization with dCRAB was performed for 100 random instances of initial and target states and the error presented in the figure is the maximum error over all random instances at given bandwidth. The fit corresponds to the error bound Eq.~\eqref{eq:bound_error} where the channel capacity is given by Eq.~\eqref{Shannon_law} and thus a scaling of the error as $\varepsilon\propto \exp(-b_1 \Delta\Omega)+b_2$ (with fit parameters $b_1$ and $b_2$) was derived from Eq.~\eqref{eq:bound_error}.
The complexity of a quantum state can also be measured by the entanglement entropy Ent$(|\psi(T)\rangle)= -\tr\left(|\psi(T)\rangle\langle\psi(T)|\log|\psi(T)\rangle\langle\psi(T)|  \right)$  (see also Sec.~\ref{sec:entanglement-generation}). It was shown how this quantity can be used as a control objective for CRAB~\cite{Caneva2011,Caneva2012} and how in turn a highly entangled state can be brought back to the ground state by means of optimal control \cite{Caneva2014} thus effectively reversing the time dynamics and entanglement generation induced by sudden quenches of the system parameters.

\subsection{Software Platforms}\label{sec:software-platforms}

The gradient-free nature of (d)CRAB makes the algorithm very versatile and allows it to be used together with virtually any available software package for the simulation of quantum systems.
The RedCRAB software package~\cite{Frank2017,Heck2018,Omran2019} is a python-based programme that incorporates this flexibility by providing many different interfaces to simulation software or experiments, e.g., by the possibility to link it to MATLAB, other python programmes, the command line, or simple file transfer on a common exchange folder with the simulation/experiment. The RedCRAB software is available from the authors on request. RedCRAB operates with a client-server scheme where the client software is installed by the user and serves as an interface between the experiment or simulation on one side and the core algorithm (running at the server) on the other side. A tailored version of the software~\cite{Rossignolo2021} is available as an extension to QUDI, a modular python suite for experiment control developed by some of the authors of this review~\cite{Binder2017}.

The software project QuTiP~\cite{Johansson2012,Johansson2013} (Quantum Toolbox in Python) that provides tools for the numerical simulation of many commonly investigated quantum systems and optimal control methods for these systems also provides an implementation of dCRAB.

The possibility to connect the dCRAB algorithm with an experiment for closed-loop optimization is described in detail in Sec.~\ref{sec:closed-loop}. For open-loop optimal control, i.e., optimization based on numerical simulations, sometimes the interplay of the simulation of time evolution and optimal control can be complicated when the optimal control algorithm does not allow for a clear separation of the optimal control code and the time evolution code (e.g., for gradient-based algorithms, where the gradient has to be calculated during numerical time evolution). Software libraries such as QuTiP and QDYN (www.qdyn-library.net) provide integrated solutions to such problems. On the other hand, CRAB does not need this integrated approach and can directly use existing software packages for numerical time evolution, which is increasingly important for more complex quantum systems. CRAB was used together with numerical simulations based on time-dependent density matrix normalization group (tDMRG)~\cite{White1992,Schollwoeck2005,DeChiara2008,Caneva2011, Caneva2011b, Doria2011}, matrix product states (MPS)~\cite{Schollwoeck2011,Frank2014,Mueller2013,Mueller2016}, tensor networks (TN)~\cite{Cui2017, Silvi2019}, multi-configurational Hartree for Bosons (MCTDHB) \cite{Brouzos2015, Alon2016} and a Krylov subspace method tailored to Rydberg atoms~\cite{Omran2019}.
This broad variety of software packages used for or together with CRAB optimization corroborates the versatility of the approach and its ready applicability to a large class of problems.

\subsection{Conclusions}
In this section we have introduced the algorithm and its different versions. We have discussed the convergence properties of the algorithms depending on the control landscape and seen that the advantageous convergence properties for unconstrained control problems can be maintained also for bandwidth limited control pulses. We have furthermore discussed the influence of limited control resources and noise on the performance of a control solution in terms of information theoretical error bounds. In the next section we show how the dCRAB algorithm can be applied to many different control problems and how the symmetries of these control problems can be exploited to reduce dimensionality (and thus complexity).

%**************************************************************
\section{Applications - toolbox}\label{sec:applications-toolbox}
In this section we discuss in more detail the control problem introduced in Sec.~\ref{sec:control-problem}  to demonstrate that dCRAB optimization is a very versatile tool that goes way beyond the simple example of state preparation using the fidelity as a control objective (Sec.~\ref{sec:control-problem-simple}) that served us as an example in the previous section. We present a broad variety of problem classes and show how to tackle them with tailored control objectives.

\subsection{Control of Quantum Many-Body Systems}\label{sec:many-body}
The (d)CRAB algorithm has been found to be a powerful tool to optimize the state preparation in quantum many-body systems as it does not depend on calculating the gradient of the control objective and can produce control pulses with very high fidelities after only a moderate number of iterations.  Especially, CRAB has been used in combination with several numerical tools simulating the time dynamics of the system, such as density matrix normalization group (DMRG) \cite{Doria2011,Caneva2011, Caneva2011b,Caneva2012}, tensor networks or matrix product states \cite{Mueller2013, Caneva2014,Frank2014,Frank2016,Mueller2016,Pichler2016,Cui2017}, or multi-configurational time-dependent Hartree for Bosons (MCTDHB) \cite{Brouzos2015}. Recently, also gradient-based optimization was applied successfully to quantum many-body systems~\cite{Jensen2020}, opening the possibility for interesting future developments of the field.

A typical control task for quantum many-body systems is state preparation of a specific state $|\zeta\rangle$.
Starting from an initial state $|\xi\rangle$, e.g., the ground state, which can be readily prepared, the control objective then reads
\begin{eqnarray}
	J(f)=F(|\psi(T)\rangle)\\
	F(|\psi(T)\rangle) = |\langle\zeta|U(T)|\xi\rangle|^2\,.
\end{eqnarray}
If we study an $L$-body quantum sytem with local dimension $d$, the total dimension of the quantum system $N=d^L$ grows exponentially with the number of constituents. For a fully controllable system the degrees of freedom for many-body state preparation are then $D_r=2N-2=2 d^L-2$~\cite{Lloyd2014,Caneva2014}; see also Sec.~\ref{sec:constrained-optimization}. However, in practice one often encounters situations where only a relatively small portion of the full Hilbert space is of interest. In fact, a very common simulation technique for many-body systems is based on tensor networks (or tDMRG) \cite{White1992, Schollwoeck2005, Schollwoeck2011, DeChiara2008},which can be used also in the framework of optimal control of many-body systems~\cite{Caneva2011b, Doria2011,Mueller2013,Frank2016,Jensen2020}. One key feature of such tensor networks is that for each possible bi-partition along the tensor network describing the quantum system the number of singular values is truncated at a fixed maximum value, the so-called bond dimension. As a consequence, if a quantum system can be efficiently simulated by this technique, the effective dimension of the system scales linearly with the number of constituents (or sites) $D_r\sim L$.

To give some examples, a W-state (a symmetric superposition of all combinations with one site in an excited state and all others in the ground state) can be described by a tensor network state with a bond dimension of $2$. Such a W-state naturally arises for example in systems of Rydberg atoms, where the strong interaction between two atoms in the Rydberg state leads to the so-called Rydberg blockade that prevents double excitation. Using a simulation based on matrix product states that allows for long-range interactions, Ref.~\cite{Mueller2013} could effectively simulate up to $30$ 3-level atoms and calculate an optimal state preparation pulse for a W-state in a thermic vapor of Rydberg atoms. With ultracold lattices of Rydberg atoms, this could be applied to the transport of a qubit over a quantum bus composed of Rydberg atoms~\cite{Mueller2016} and to symmetric superpositions of different Fock states and the GHZ state~\cite{Cui2017}. Also the highly entangled GHZ state is described by tensor networks of bond dimension $2$. This allowed the creation of a $20$-qubit Schrödinger cat with QOC-enhanced state preparation pulses on a lattice of ultracold Rydberg atoms~\cite{Omran2019}.

In some cases, also a mean-field description of the many-body system is possible. In an experiment with a BEC of Rubidium atoms on a chip a state transfer between the motional ground state and the first excited motional state was optimized by CRAB~\cite{Frank2016}. The system was simulated by a numerical solution of the Gross-Pitaevski equation and the optimal control algorithm could design a modulation of the trap potential that lead to an experimental fidelity of the state preparation of more than $99\,\%$. On the same system a similar process was implemented to realize a Ramsey interferometer~\cite{Frank2014} (see Secs.~\ref{sec:enhanced-measurements},~\ref{sec:BEC}).

Another way to decrease the control complexity -- if available -- is to start from an intuitive or adiabatic solution of the control problem. This is often the case when the initial and final state can be expressed as the ground states of the same Hamiltonian with different parameters. An example is the preparation of a Mott insulator state with cold atoms in an optical lattice~\cite{Doria2011,Rosi2013,Frank2016}. In this case, the ground state of the Mott insulator phase (with exactly one atom per site) could be reached from the ground state of the superfluid phase by carefully ramping up the lattice depth. In the experiment, the CRAB-engineered ramp was three times faster than the adiabatic solution at comparable fidelity. 
Recently, further improvement was obtained numerically by introducing a gradient algorithm suitable to tackle quantum many-body systems~\cite{Jensen2020}.  The speed-up that can be obtained by QOC compared to adiabatic solutions can be analyzed by considering the spectral gap $\Delta$ at the position of the phase transition~\cite{Caneva2011b}. For adiabatic solutions, the pulse operation time has to be much larger than the inverse spectral gap: $T\gg \Delta^{-1}$. Otherwise, the adiabatic solution will lead to unwanted excitations of the sytem~\cite{Sachdev1999,Zurek2005,Polkovnikov2008}.
For QOC solutions, instead, the QSL is given by $T_{\mathrm{QSL}}\approx \frac{\pi}{\Delta}$ and this relation was confirmed numerically for the Lipkin-Meshkov-Glick model for a phase transition from paramagnetic to ferromagnetic phase, as well as for the Grover search algorithm and the two-level Landau-Zener model~\cite{Caneva2011b}.

\begin{figure}[ht]
	\centering
	\includegraphics[width=0.48\textwidth]{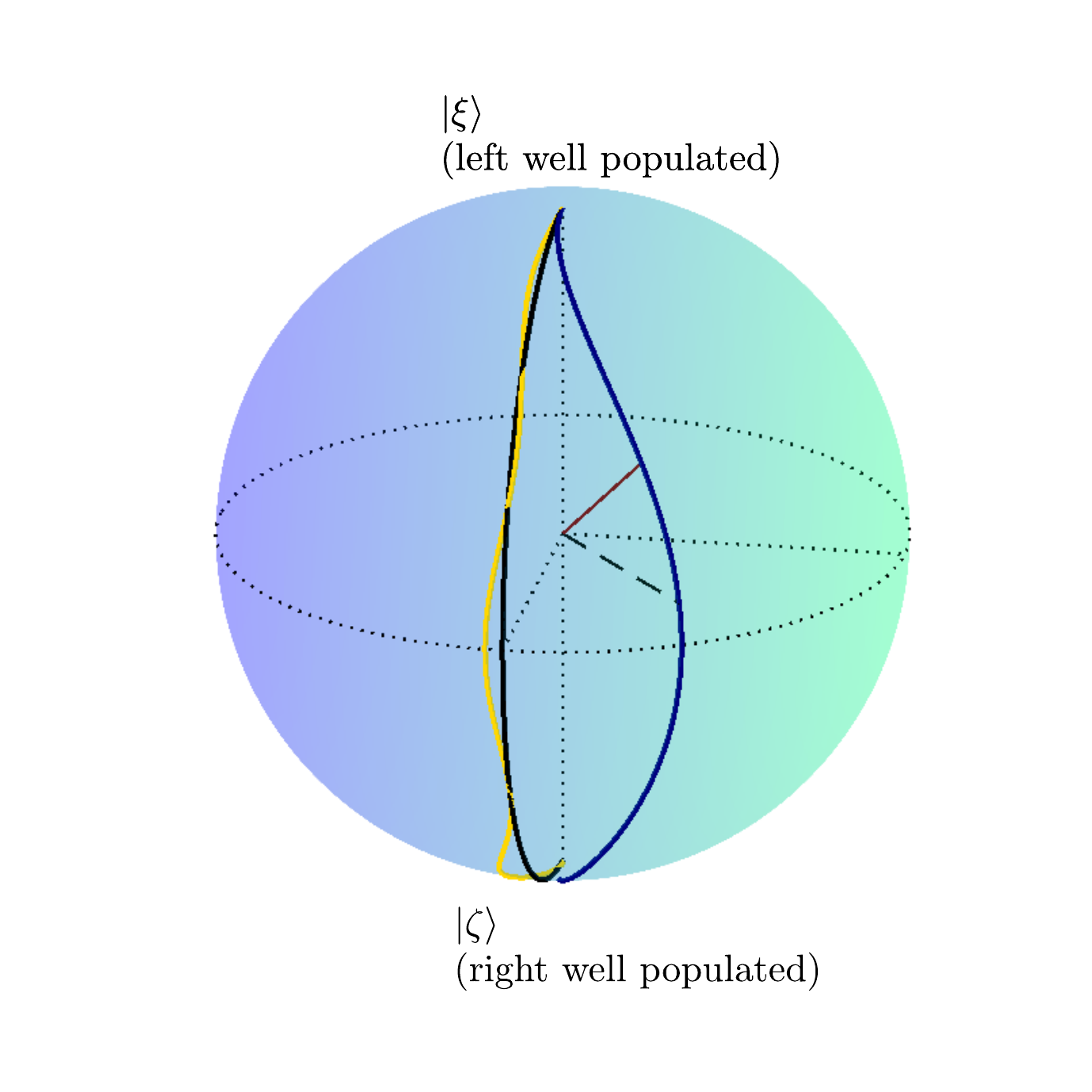}
	\caption{Optimal control path at the QSL: A BEC of interacting atoms is initially trapped in one site of a double well potential. The figure shows the evolution of the system as a path in the Bloch sphere given by the initial and final state. Under free evolution tunneling occurs, while non-linear effects lead to a deviation of the path (blue line) from the geodesic (black line) that connects the initial state $|\xi\rangle$ with the final state $|\zeta\rangle$.  By properly designed control pulses the state transfer is achieved with high fidelity even when the operation time approaches the QSL. The path of the controlled evolution (yellow) then approaches the geodesic. The data for the figure is taken from Ref.~\cite{Brouzos2015}.}
	\label{fig:geodesic}
\end{figure}

Also the dimension of the effective Hilbert space of the many-body system can be investigated by a closer look at the QSL (see Sec.~\ref{sec:constrained-optimization}) in the context of quantum many-body systems. In Ref.~\cite{Brouzos2015} a BEC of interacting atoms (see Sec.~\ref{sec:BEC})  in a double well potential (or Bosonic Josephson Junction) was studied under three different approximations that alter the dimension of the system model: a) the two-mode approximation (given by two eigenstates of the Gross-Pitaevski equation corresponding to the two wells) b) the Gross-Pitaevski equation mean-field approximation and c) the full quantum-many body Schrödinger equation solved by MCTDHB. Then, control strategies were examined for a transfer of the atoms from one well to the other. The QSL is given by the coupling that drives the tunneling through the barrier. However, non-linear effects push the system in a different direction and this effect has to be mitigated by suitable control pulses. When the pulse operation time approached the QSL, the optimal transfer in all three scenarios was a transfer along the geodesic of the Bloch sphere given by the initial and target state (see Fig.~\ref{fig:geodesic}).

We conclude by remarking that, for many-body systems, understanding the control complexity is a crucial step toward achieving control over the system and that wherever possible, one has to try to decrease the control complexity by restricting the problem to the most important dimensions of the system (as in the case of tensor networks or mean-field approaches), by performing a local search around an intuitive solution and/or by introducing additional degrees of freedom in the control objective. This latter approach is further described in Secs.~\ref{sec:entanglement-generation}~and~\ref{sec:enhanced-measurements} and was first demonstrated with CRAB for the case of entanglement generation~\cite{Caneva2012}.

\subsection{Closed-loop Optimization}\label{sec:closed-loop}
In this section we explain how the dCRAB algorithm can be used for closed-loop optimization of experiments.
By closed-loop optimization we intend an optimization of the control pulse in a (classical) feedback loop involving the system (i.e., experiment or quantum device), where the feedback from the system consists of the measurement of the control objective at final time.
A closed-loop optimization is thus possible, if the experiment itself (or more precisely a quantum measurement at the end of the time dynamics) is used to evaluate the control objective for each iteration of the optimization \cite{Rosi2013}. This is possible even if the optimization algorithm runs on a server that has only remote access to an experiment located in a different lab~\cite{Frank2017} or country~\cite{Heck2018}.

The key advantage of closed-loop optimal control (compared to open-loop strategies) is that the description of the experiment in a simulation has always limited precision, since not all aspects and noise sources of the experiment are fully known or they cannot be taken into account in an efficient simulation, especially for large systems (or) in more than one dimension~\cite{Vidal2004}. By evaluating the control objective directly from the measurement outcomes of the experiment, this imprecision in the simulation can be circumvented. Still, also the experimental evaluation of the control objective is time consuming, especially when high precision is needed and thus many individual quantum measurements are required~\cite{Paris2004,Pezze2018}. Also, in such cases, no gradient information is available, which is a major reason to operate in a chopped basis with direct optimization, unless one wants to approximate the gradient by many additional evaluations of the control objective~\cite{Brif2010}. As in the case of many-body optimization (see Sec.~\ref{sec:many-body}), it is helpful to have already some prior knowledge or good guess about the optimal pulse shape. Such an initial guess can come from the adiabatic solution of a many-body phase transition~\cite{Rosi2013}, from a numerical simulation of the system~\cite{Frank2017}, or from human intuition (in \cite{Heck2018} this human intuition was supported by so-called citizens scientists playing a game-version of the quantum control problem; see www.scienceathome.org).
\begin{figure}[ht]
	\centering
	\includegraphics[width=0.7\textwidth]{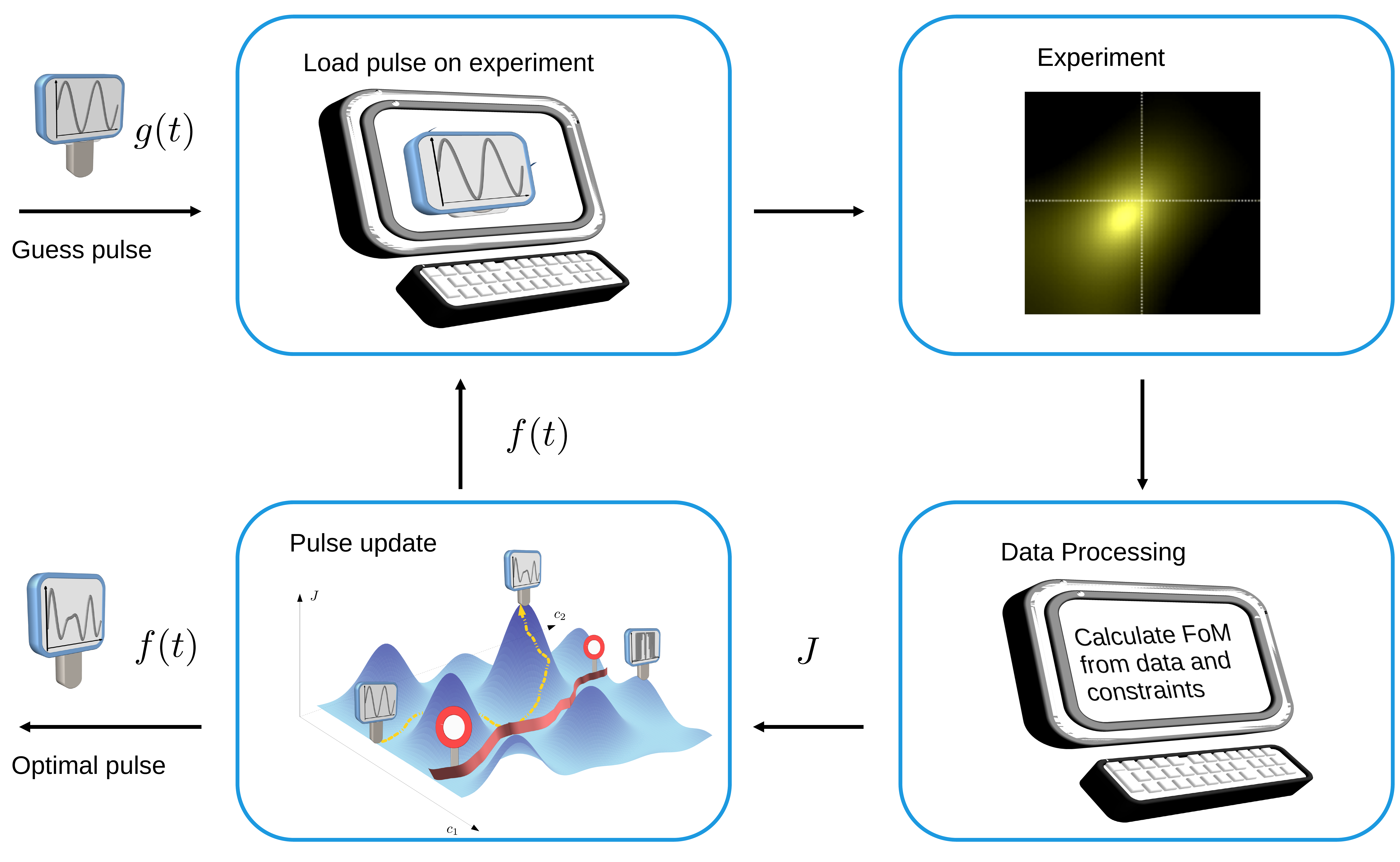}
	\caption{Closed-loop Control: A guess pulse $g(t)$ is sent to the experiment. Then, the figure of merit (FOM; control objective) $J$ is evaluated by processing the experimental data and sent back to the algorithm. The control algorithm calculates a new pulse and the loop is repeated. The guess pulse $g(t)$ is iteratively transformed into the optimized pulse $f(t)$.}
	\label{fig:closed-loop-control}
\end{figure}
Fig.\,\ref{fig:closed-loop-control} gives a schematic overview on the procedure: a guess pulse $g(t)$ is loaded to the experiment via the lab software and the control electronics. The experiment is then performed several times with this control pulse to accumulate enough data for the required precision. Afterwards, the collected data from the experiment is processed to calculate the value $J$ of the control objective associated with the trial control pulse. This value $J$ (and possibly an estimate of the error attached to this value) is then used by the dCRAB algorithm to update the control pulse, e.g., according to Eq.~\eqref{eq:CRAB-guess}. Finally, a new pulse $f(t)$ is fed back to the lab software for a new iteration of the optimization algorithm.

Concluding, this technique can be used to calibrate existing control solutions (e.g., from open-loop optimization) to the control electronics, or to account for effects coming from the limited accuracy of the simulation due to perturbative models, additional effects from the system-environment interaction or external fields. If the experimental evaluation is fast enough, the closed-loop optimization can also completely substitute the open-loop simulation.

\subsection{Entanglement Generation}\label{sec:entanglement-generation}
Entanglement~\cite{Horodecki2009} is one of the most interesting characteristics of a quantum system and an important resource in quantum metrology~\cite{Paris2004,Pezze2018}, quantum sensing~\cite{Degen2017}, quantum communication~\cite{Duan2001} and quantum computation/simulation~\cite{Nielsen2010}. Here, we show different ways to optimize control pulses for the generation of entanglement.

One obvious way to generate entanglement in a system is to prepare a specific entangled state, e.g., a GHZ state
\begin{equation}
	|\zeta\rangle = \frac{|0\dots 0\rangle + |1\dots 1\rangle}{\sqrt{2}}.
\end{equation}
Starting from an initial state $|\xi\rangle$, e.g., the ground state, that can be readily prepared, the control objective reads
\begin{eqnarray}
	J(f)=F(|\psi(T)\rangle)\\
	|\psi(T)\rangle=\Gamma_f(|\xi\rangle)\\
	F(|\psi(T)\rangle) = |\langle\zeta|\psi(T)\rangle|^2\,.
\end{eqnarray}
This approach is very successful when the final state $|\zeta\rangle$ is chosen in such a way that it fits the symmetries of the system. For example with Rydberg atoms a W-state \cite{Mueller2013} and GHZ state \cite{Omran2019} have been achieved through CRAB/dCRAB optimization. The drawback of this state preparation, however, is that it fixes $D_r=2N-2$ degrees of freedom, when in reality the amount of entanglement itself is just one parameter (let us call it Ent$(|\psi(T)\rangle)$, the (generic) entanglement measure). In principle any entanglement measure \cite{Horodecki2009} can be used as control objective. Choosing  Ent$(|\psi(T)\rangle)$ as control objective fixes only one degree of freedom of the final state ($D_r=1$), and thus potentially less control parameters are needed to maximize the control objective. Indeed, numerically entanglement could be generated for a very large system: for a many-body system (Lipkin-Meshkov-Glick model) using the entanglement entropy Ent$(|\psi(T)\rangle)= -\tr\left(|\psi(T)\rangle\langle\psi(T)|\log|\psi(T)\rangle\langle\psi(T)|  \right)$  as control objective CRAB was able to saturate the entanglement for up to $L=100$ qubits~\cite{Caneva2012}. Note, that the preparation of a specific entangled state according to the ``$2N-2$"-rule (see Secs.~\ref{sec:constrained-optimization},~\ref{sec:many-body})  can be virtually impossible for such a large system.

Yet another possibility to generate entanglement is the application of an entangling gate onto a suitable initial state. This is discussed in the following section.

\subsection{Quantum Gates}\label{sec:quantum_gates}

In this section we present different control objectives that allow to optimize quantum gate operations. For the purpose of optimal control a quantum gate $V$ is achieved when the time evolution leads to a unitary operation $U$ at final time $T$ that is equivalent to $V$ up to some symmetries. These symmetries can be for example the global phase or local transformations. These symmetries have to be encoded in the control objective, usually through the gate fidelity. Refs.~\cite{Palao2003,Rembold2020} present the three most common choices:
\begin{eqnarray}
	F_{Re}=\frac{1}{N}\re \tr U^\dagger V\label{eq:F_Re},\\
	F_{sm}=\frac{1}{N^2}\vert \tr U^\dagger V\vert^2\label{eq:F_sm},\\
	F_{ss}=\frac{1}{N^2}\sum_{k,l=1}^N \vert u_{kl}^*v_{kl} \vert^2,
\end{eqnarray}
where $F_{Re}=1$ requires $U=V$, while $F_{sm}=1$ allows for a different global phase of $U$ and $V$, and $F_{ss}$ disregards the phases of every matrix element ($u_{kl}$ and $v_{kl}$ denote the matrix elements of the gates $U$ and $V$, respectively). These fidelity functions can readily be optimized by chopped random basis optimization, as done for example with a few-parameter control for a two-qubit Rydberg atom phase gate~\cite{Mueller2014} and for a single-qubit gate with NV centers~\cite{Scheuer2014,Frank2017}.

If we want to allow for more degrees of freedom in the target gate, the control objective becomes more complicated and a comprehensive study was performed for two-qubit gates \cite{Mueller2011,Watts2015,Goerz2015,Goerz2019} (see also Refs.~\cite{Zhang2003a,Zhang2003b,Zhang2004a,Zhang2004b,Makhlin2002} for the theory of algebraic decomposition of two-qubit gates). The key idea is that a gate $U\in SU(4)$ can be decomposed into its non-local (two-qubit) content $A$ and local operations $k_{1},k_{2}\in SU(2)\otimes SU(2)$ via
\begin{eqnarray}\label{eq:Cartan}
	U=k_{1}Ak_{2},\quad\text{where}\\
	A=\mathrm{exp}\left(\frac{i}{2}(c_{U,1}\sigma_x\otimes\sigma_x +c_{U,2}\sigma_y\otimes\sigma_y + c_{U,3} \sigma_z\otimes\sigma_z)\right)\,.
\end{eqnarray}
Here, $\sigma_x$, $\sigma_y$, and $\sigma_z$ are the Pauli matrices and the coefficients $c_{U,1},c_{U,2},c_{U,3}\in \RR$ characterize all global characteristics of the gate (like its entangling power or universality) and can be calculated from the matrix representation of $U$ in the canonical basis.
If we now consider a target gate $V$ with its non-local content given by the coefficients $c_{V,1},c_{V,2},c_{V,3}$ and the optimized gate $U$ given by $c_{U,1},c_{U,2},c_{U,3}$, we can modify the fidelity Eq.~\eqref{eq:F_Re} by including the symmetry of local (single qubit) transformations:
\begin{eqnarray}
	\label{eq:fidelity-unitary}F^{NL}_{Re}(U,V)&=&\max_{k_1,k_2}\frac{1}{4}\re \tr U^\dagger k_1 V k_2 \nonumber\\
	&=&\cos\frac{\Delta c_1}{2}\cos\frac{\Delta c_2}{2}\cos\frac{\Delta c_3}{2}\,,
\end{eqnarray}
with $\Delta c_i= c_{U,i}-c_{V,i}$ ($i\in \{1,2,3\}$). If the result of the optimization $U$ is not necessarily unitary, the control objective can be modified by help of $\tilde{U}$, the closest unitary to $U$:
\begin{eqnarray}
	J=F^{NL}_{Re}(\tilde{U},V)-\Vert U-\tilde{U}\Vert\,.
\end{eqnarray}
Sometimes, also the exact non-local content of the gate is not important and one wishes instead only to obtain a gate with maximal entangling power. Such a gate is called a perfect entangler and is characterized by its power to transform a product state into a maximally entangled state (e.g., a Bell state). This perfect entangler property is given by geometrical equations for the coefficients $c_{U,1},c_{U,2},c_{U,3}$ \cite{Watts2015,Zhang2003a}, namely by the inequalities
\begin{align}
	&c_{U,1}+c_{U,2}\geq \frac\pi 2\,\nonumber\\
	&c_{U,1}-c_{U,2}\leq \frac\pi 2\,\nonumber\\
	&c_{U,2}+c_{U,3}\leq \frac\pi 2\,.
\end{align}
From there, a fidelity function can be formalized as
\begin{eqnarray}\label{eq:fidelity-perfent-unitary}
	F^{PE}_{Re}=\begin{cases}
		1 & U \text{ is a perfect entangler}\\
		\max (\cos^2\frac{c_{U,1} + c_{U,2}- \frac{\pi}{2}}{4},& \nonumber\\
		\qquad\;   \cos^2\frac{c_{U,1} - c_{U,2}- \frac{\pi}{2}}{4},& \nonumber\\
		\qquad\;   \cos^2\frac{c_{U,2} + c_{U,3}- \frac{\pi}{2}}{4}) & \text{ otherwise.}
	\end{cases}
\end{eqnarray}
This method can be used also to investigate which gates are accessible with a given physical system (depending on the controllability of the system and possible leakage to auxiliary levels or spontaneous emission)~\cite{Goerz2015}.

\begin{figure}
	\subfigure{\includegraphics[width=0.5\textwidth]{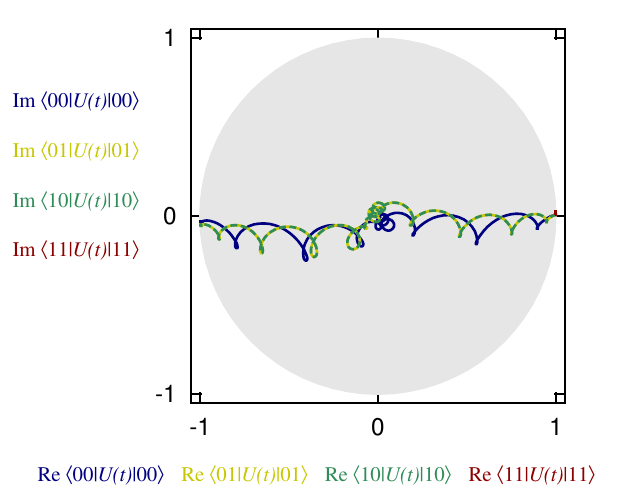}}\hfill
	\subfigure{\includegraphics[width=0.5\textwidth]{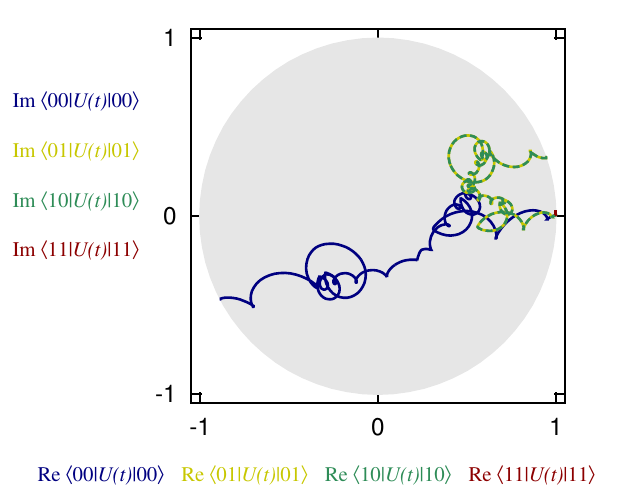}}
	\caption{The left panel shows how each basis state acquires a phase according to the desired target gate $V=\mathrm{diag}(-1,-1,-1,1)$. The right panel shows the time evolution of the phases and shows that the entangling phase $\phi_{1} - \phi_{2} - \phi_{3} + \phi_{4}\stackrel{!}{=}\pi$ is distributed over all four basis states according to what fits best to the Hamiltonian of the system. This is an effect of the control objective incorporating local invariants. The data for the figure ist taken from~\cite{Mueller2011}.}
	\label{fig:RydbergLI-dynamicsnoloss}
\end{figure}

An illustrative example is given by the case of phase gates and additional single-qubit phases.
We consider two diagonal gates $U=\mathrm{diag}( \mathrm{e}^{i\phi_{U,1}}, \mathrm{e}^{i\phi_{U,2}}, \mathrm{e}^{i\phi_{U,3}}, \mathrm{e}^{i\phi_{U,4}})$ and $V=\mathrm{diag}( \mathrm{e}^{i\phi_{V,1}}, \mathrm{e}^{i\phi_{V,2}}, \mathrm{e}^{i\phi_{V,3}}, \mathrm{e}^{i\phi_{V,4}})$ and set $\Delta\phi_i := \phi_{U,i} - \phi_{V,i}$.
The gate fidelity with which $U$ approximates $V$ then reads
\begin{eqnarray}
	F_{Re}=\frac{1}{4}\re\sum_{i=1}^4\mathrm{e}^{-i\Delta\phi_i}= \frac{1}{4}\sum_{i=1}^4\cos\Delta\phi_i \,\text{.}
\end{eqnarray}
Single qubit phase operations are of the form $\vert\alpha\rangle_j \rightarrow \mathrm{e}^{i s_j^\alpha}\vert\alpha\rangle_j$ \cite{Calarco2001} ($s_j^\alpha$ is the rotation angle, and $\alpha=0,1$ stands for the single qubit basis $\vert 0\rangle$ and $\vert 1\rangle$ and $j=1,2$ denotes the first and the second qubit.). The application of such single qubit phase operations leaves the phase difference $\Delta\phi:=\Delta\phi_1 - \Delta\phi_2 - \Delta\phi_3 + \Delta\phi_4$ unchanged. Note that these single qubit phase operations are exactly the local transformations under which $U$ and $V$ remain diagonal.
If we now maximize $F_{Re}$ as a function of  the $\Delta\phi_i$ under the constraint
$\Delta\phi=\Delta\phi_1 - \Delta\phi_2 - \Delta\phi_3 + \Delta\phi_4 = const.$ we obtain
\begin{eqnarray}
	\Delta\phi_{1}=-\Delta\phi_{2}=-\Delta\phi_{3}=\Delta\phi_{4}=\Delta\phi/4\,,\\
	F^{NL}_{Re}=\cos\frac{\Delta\phi}{4}\label{eq:fidelity_localphase}\,\text{.}
\end{eqnarray}
The effect of this symmetry can be seen in Fig.~\ref{fig:RydbergLI-dynamicsnoloss}. It illustrates for a Rydberg phase gate~\cite{Mueller2011} how the use of Eq.~\eqref{eq:F_sm} as control objective as compared to Eq.~\eqref{eq:fidelity_localphase} influences the phase evolution of the basis states under $U(t)$ for the same target gate $V=\mathrm{diag}(-1,-1,-1,1)$.

The Rydberg phase gate can be used directly to generate entanglement. For example~\cite{Mueller2016}, if we prepare the two qubits in the ground state $|00\rangle$ and drive it to the superposition state
\begin{eqnarray}
	\frac{|0\rangle-i|1\rangle}{\sqrt{2}}\otimes \frac{|0\rangle-i|1\rangle}{\sqrt{2}}
\end{eqnarray}
by a local $3\pi/2$-pulse (or $-\pi/2$) on $\sigma_x$ for both qubits, and then apply the phase gate $V=\mathrm{diag}(1,-1,-1,-1)$, we obtain the entangled state
\begin{eqnarray}
	\frac{|00\rangle+i|01\rangle +i|10\rangle +|11\rangle}{2}\,.
\end{eqnarray}
Another local $\pi/2$-pulse on $\sigma_x$ for the second qubit finally yields the Bell state
\begin{eqnarray}
	\frac{|00\rangle +|11\rangle}{\sqrt{2}}\,.
\end{eqnarray}
Summarizing, we have presented different ways to design control objectives for the optimization of quantum gates depending on the degrees of freedom that one wants to include in the optimization. For the example of a two-qubit gate, fixing all degrees of freedom prior to optimization leads to a 16-dimensional optimization, while through various symmetries this can be reduced to one parameter in the case of the perfect entangler optimization. On the one hand, this can be advantageous in the presence of noise or when only partial control over the system is available. On the other hand, the reduction of degrees of freedom simplifies the control problem in all cases and requires less complex control pulses (see Sec.~\ref{sec:constrained-optimization}).

\subsection{Enhanced Measurements}\label{sec:enhanced-measurements}
In various occasions, precise quantum measurements highly depend on non-trivial unitary operations\,\cite{Pezze2018,Degen2017}, e.g., when the desired quantity is not directly accessible or when the precision can be tuned by preparing non-trivial states. In particular, (d)CRAB has been used to map motional states on spin states in trapped ions \cite{Mueller2015}, to enhance atom interferometry with a BEC \cite{Frank2014,Lovecchio2016}, to prepare a squeezed spin state \cite{Pichler2016}, to demonstrate an enhanced Casimir effect \cite{Hoeb2017} and to construct optimal frequency filter functions in noise spectroscopy \cite{Mueller2018}.

For a trapped ion~\cite{Leibfried2003,Huber2008,Monz2011, Walther2012, Casanova2011, Singer2010, Zhang2018} the quantum state is given by both the motional and the spin state of the ion. However, the read-out can detect only whether the ion is in a given spin state. As a consequence, the motional state of the ion can be measured only indirectly. In Ref. \cite{Mueller2015} CRAB was used to map the population of a given motional state onto one of the two spin-qubit states and the population of all other motional states onto the other spin-qubit state. In this way, a measurement of the spin population could give the occupation of the respective motional state and the procedure could also be used to prepare motional Fock states. The preparation and readout of motional Fock states can also be achieved through repetitive spin state measurements with partial selectivity to the motional state, as proposed in Ref.~\cite{Huber2008}. In this reference, the technique was used to propose the demonstration of the quantum Jarzinsky equation with a trapped ion.

The dynamical Casimir effect can be observed when a qubit is placed in a cavity and its frequency is rapidly moved in and out of resonance with the cavity \cite{Hoeb2017}. Then, photons are generated in the cavity out of the vacuum. This effect can play a role in superconducting circuits \cite{Wilson2011,Lahteenmaki2013} and in Ref.~\cite{Hoeb2017} it could be shown that QOC (and in particular the dCRAB algorithm) can be used to modulate the qubit resonance in such a way that the visibility of the dynamical Casimir effect is enhanced by an order of magnitude with respect to dynamical decoupling and by another order of magnitude compared to the basis scenario of a simple frequency sweep.

An atomic interferometer was studied with a BEC of Rubidium atoms on a chip in Refs.~\cite{Frank2014,Lovecchio2016}. In Ref.~\cite{Frank2014} a Ramsey interferometer was realized using the motional ground state and the first excited motional state. A temporal modulation in the trap potential was engineered by CRAB to perform the Ramsey-$\pi/2$-pulses between the two states, where the control objective was the contrast between the measurement outcomes. This contrast was increased to over $90\,\%$ and allowed to measure the contribution of the interaction energy in the phase evolution of the excited state.  In Ref.~\cite{Lovecchio2016}, instead, CRAB optimal control enabled the preparation of a superposition of hyperfine states with maximized energy splitting, and these states could then be used for Ramsey interferometry, showing a four-fold improvement of the sensitivity of the atomic interferometer.

Ref.~\cite{Pichler2016} employs QOC to prepare squeezed spin states. Squeezed states allow to improve the precision of a quantum observable by squeezing its variation at the expense of stretching the variation of a conjugate observable while maintaining the Heisenberg uncertainty relation \cite{Pezze2018}. In reference~\cite{Pichler2016}, optimally controlled state preparation pulses allowed to speed up the preparation time, where the speed-up scaled also favorably with the system size compared to the standard protocol (based on adiabatic one-axis twisting). The optimal control solutions turned out to be more robust against noise (e.g. spin-spin interaction fluctuations in a many-body system) during state preparation, and as a result promise to increase the achievable squeezing and thus the measurement precision.

Finally, in noise specroscopy the decoherence function $\chi(T)$ of a two-level quantum probe can be written in the \emph{universal form} \cite{Kofman2000,Kofman2001,Degen2017,Szankowski2017}
\begin{eqnarray}
	\chi(T)=\int_{0}^\infty \mathcal{F}(\omega)S(\omega) d\omega\quad\text{with}\\
	\mathcal{F}(\omega)=\left|\int_0^T y(t) e^{-i\omega t} dt\right|^2\,,
\end{eqnarray}
where $S(\omega)$ is the spectrum of the noise (or signal that one wants to encode into the decoherence function), $\mathcal{F}(\omega)$ is the so-called frequency filter function and $y(t)$ is the pulse modulation function. The pulse modulation function can be for example a function $y\in\{-1,1\}$ that flips sign each time a $\pi$-pulse is applied to the quantum probe, but it can also be a continuous function. By engineering the pulse modulation function $y(t)$ one can shape the decoherence function and/or encode specific information in it. Optimization methods have been employed to minimize the decoherence under the constraint of limited pulse energy~\cite{Gordon2008} (by variational principle) as well as to shape $\mathcal{F}(\omega)$ to increase the sensitivity (maximize the Fisher information) with respect to a specific (deterministic) temporal field while minimizing the effect of noise \cite{Poggiali2018} (through direct optimization of the pulse spacings of a dynamical decoupling sequence), and also in order to maximize the Fisher information with respect to a signal of given spectrum \cite{Mueller2018} using dCRAB to optimize $y(t)$.

Summarizing, we have presented examples for the use of QOC to measure properties of quantum systems with high precision by either preparing the necessary superposition states, e.g., for interferometry, or by mapping the quantity of interest onto a state that can be readily measured. Such a mapping is possible for composite systems, where, e.g., in a spin-boson system bosonic properties are mapped on the spin state or vice versa,  as well as for system-environment interactions, where properties of the environment are mapped onto the state of the system.

\subsection{Opimization of open system dynamics}

Open quantum systems offer a special challenge to QOC~\cite{Schmidt2011,Mukherjee2013,Hoyer2014,Kallush2014,Pawela2015,Mukherjee2015,Reich2015,Lovecchio2016,Mueller2020,Koch2016} since many of the conventional statements about controllability and control landscapes hold just for closed systems and instead for open quantum systems the set of reachable states is usually restricted to a subset of the whole Hilbert space~\cite{Koch2016}. Yet, the degree of controllability and the influence of the environment can also reveal information on the environment~\cite{Degen2017}.

While every real quantum system is an open system since it can never be perfectly shielded from its environment, in many cases the environment can be treated as a small perturbation of the unitary system dynamics.
As discussed already in Sec.~\ref{sec:constrained-optimization}, additional control resources can be used to reduce the effect of the non-unitary contributions to the open system dynamics. While this does not work in all cases and also with infinite resources usually not every state can be reached~\cite{Koch2016}, there is numerical evidence of  the control error scaling according to Eq.~\eqref{eq:bound_error} that reflects the information damping given by the noise (or decoherence)~\cite{Mueller2020}. In particular, for non-Markovian systems, where the noise is correlated and high-frequency noise is negligible, a control bandwidth that exceeds the correlation time can lead to very small control errors.

In some cases, however, the influence of the environment on the system dynamics is the main focus of the investigation or can even be used as a control resource.
In \cite{Caruso2012,Hoyer2014} the evolution of electronic excitations in the Fenna-Matthews-Olson complex \cite{Olson2004} -- a biological pigment-protein complex involved in the early steps of bacterial photosynthesis -- was studied. The coherent dynamics of these excitations are disturbed by the decoherence and dissipative effects of the protein environment. In \cite{Caruso2012} different initial states (bright and dark states) were prepared by CRAB and then the energy transfer through the complex into the reaction center (sink) was analyzed for these initial states in terms of the role of both coherent dynamics and decoherence. Ref.~\cite{Hoyer2014} investigated the controllability of the system through modulated pulses: in particular the ability of the control to localize the population in one site or in a specific exciton state. Control over the coherent part of the system could then pave the way to sensing of the spectrum of the protein environment.

In quantum thermodynamics, open system dynamics can be studied with regard to the (ir)reversibility of processes and work performed by the system. Ref.~\cite{Rach2016} showed how the irreversibility introduced by a fast change in a system parameter can be substantially reduced by CRAB optimization compared to a linear ramp or a sudden quench. At the same time the work produced by the change of the parameter was almost constant among the three protocols. Irreversible transformations with a substantial work production are key building blocks for efficient quantum cycles that produce work with only small amounts of quantum entropy production.

Ref.~\cite{Mukherjee2013} studied the role of optimal control in speeding up or slowing down relaxation processes. In a two-level system a thermal bath pushed the system toward a fixed point. Under certain conditions the process could be slowed down or enhanced by QOC (thus switching between the Zeno and anti-Zeno regime~\cite{Kofman2001}), where CRAB was used to calculate QOC solutions under constraints. The problem was also studied for non-Markovian noise to understand the interplay of memory effects and unitary control~\cite{Mukherjee2015}.

\subsection{Conclusions}
In this section we have presented the necessary QOC toolbox to address a large variety of systems and control problems. In the next section, we will discuss the most common quantum technology platforms with a focus on how this toolbox can be used on the specific platforms to increase the performance of quantum technology tasks or study fundamental physics phenomena.

%**************************************************************
\section{Applications on different platforms}\label{sec:applications-per-platform}
The (d)CRAB algorithm has been successfully applied to control problems on numerous different physical platforms. 
In this section, we present  these platforms with a focus on the specific challenges that 
have been overcome with the (d)CRAB optimization. 

\subsection{Cold atoms}\label{sec:cold_atoms}

Atomic physics has been remarkably advanced by the groundbreaking techniques of laser cooling 
and trapping at sub-milli-Kelvin experiments that started almost forty years ago~\cite{Rushton2014}. Significant theoretical and experimental progresses in many-body physics phenomena ranging from the Mott-Hubbard transition in optical lattices to strongly interacting gases in one and two dimensions have been made possible and observed in dilute, ultracold gases~\cite{Bloch2008}. Cold atoms provide an ideal platform for quantum simulation that allows to probe quantum magnetism, to detect 
topological quantum materials, and to investigate quantum systems with controlled long-range 
interactions~\cite{Gross2017,Diehl2008}. Furthermore, cold atoms find their applications in various 
technologies, such as precise inertial sensing based on light-pulse atom interferometry at long 
interrogation times~\cite{Dickerson2013}, ultra-stable atomic clocks using spin-polarized, ultra-cold 
atomic ytterbium~\cite{Hinkley2013} or the realization of portable gravimeters~\cite{deAngelis2011}. The unprecedented precision, sensitivity, and stability of these 
technologies in return lead to advancements in the testing of physics foundations, for instance the equivalence principle~\cite{Dimopoulos2007}, and the variations of fundamental constants~\cite{Marion2003}.

Optical lattices with ultracold atoms were among the first platforms studied by the CRAB algorithm and CRAB optimization allowed to find a way to substantially speed up the time required to transform a superfluid gas into a Mott insulator state~\cite{Doria2011}. The system, described by the Bose-Hubbard model, was numerically simulated with time-dependent density matrix renormalization group~(tDMRG). A quantum phase transition occurs due to changes in the optical lattice depth, which can be controlled and optimized. 
Modulations of the lattice depth were performed by the CRAB algorithm for various system sizes, up to~40 lattice sites, to drive the system into the final Mott insulator state with a faster process time 
and a higher state fidelity than the adiabatic guess pulse, and to mitigate the effects of atom number fluctuations, different initial fillings of the optical lattice, and pulse distortions. The numerical simulations showed that the superfluid gas can be transformed into the Mott insulator state with an about two orders of magnitude faster pulse than in earlier experiments~\cite{Stoeferle2004} relying on adiabatic ramps. The work developed fundamental techniques and optimization recipes for subsequent experimental realizations~\cite{Rosi2013,Frank2016,Heck2018}.

Indeed, as mentioned above, when moving to an experiment, CRAB optimization can also be performed during the experimental repetitions of the measurement process, i.e., performing a closed-loop optimization of the process. The work by Rosi,~et.~al.~\cite{Rosi2013}, experimentally demonstrated for the first time the capability of the CRAB algorithm to perform such closed-loop optimization of two key dynamical processes that are often found in various experiments with cold atoms. In this experiment, a loading process of a Bose-Einstein condensate of Rubidium-87 atoms in a two-dimensional optical lattice to produce an array of one-dimensional gases, and the driving of the gas across the quantum phase transition from the superfluid to the Mott insulator phase in a three-dimensional lattice were optimized. In both processes the CRAB method was used to modulate the lattice potential depth such that the residual motional excitation of the atoms in the final state was minimized. For each optimized lattice potential depth the final thermal fraction of the sample was measured and compared with the initial thermal fraction. The ratio between these fractions was used as the figure of merit (FOM) to be minimized.
The optimization was repeated until it converged or reached a predetermined precision. The loading process, performed as a test for the closed-loop CRAB method, was optimized over a restricted class of functions, namely loading ramps of exponential shape with different durations and time constants. The final FOM measured for the optimized loading was approximately~15\% lower than that for the quasiadiabatic loading.
Moreover, the optimized loading ramps were three times faster than the adiabatic dynamics. The full closed-loop CRAB method was subsequently performed to optimally drive the quantum phase transition from the superfluid to the Mott insulator phase. The optimization used two CRAB frequency components, therefore it searched for a minimum in a four-dimensional parameter space. Similar to the loading experiment, the phase transition driven by the optimized ramp was roughly three times faster than the adiabatic one with an improved final FOM of approximately~9\%. 

Ref.~\cite{Frank2016} studied the crossing of the phase transition from the superfluid phase to the Mott insulator phase at the quantum speed limit. In the experiment an optical lattice of 32 sites was prepared in two parallel tubes containing on average 16 Rubidium-87 atoms each. The dynamics of the finite-size one-dimensional crossover of the atoms in the optical lattice was simulated by tDMRG and optimized by CRAB via open-loop optimization following the techniques developed in~\cite{Doria2011}.
The optimization resulted in a non-adiabatic fast steering across the one dimensional superfluid-Mott insulator crossover with a speed of a factor ten faster than the adiabatic protocol, and without additional final state distortions.
The CRAB algorithm found the optimal ramp of the lattice depth for various transformation times by minimizing the rescaled average variance of the site occupation in the trap centre, where the effect of the harmonic potential is expected to be negligible and the Mott insulator state can be observed. The experimentally found quantum speed limit of approximately 12~ms corresponds to the independent theoretical prediction based on the minimal energy gap of the system, which was numerically calculated for a system of~16~particles. The optimal steering process was demonstrated to be robust against the total atom number fluctuations thus protecting the state from the influence of a realistic non-zero initial temperature of the system.

Together with the feasibility of miniaturizing magneto-optical traps~\cite{Rushton2014}, the (d)CRAB optimization method applied to the cold atoms platform will continue to lead to high-fidelity ultra-cold quantum technology devices as demonstrated also very recently by its contribution to transporting an atomic wavepacket over a distance 15 times its size at the quantum speed limit by engineering the optimal modulation of the trap potential to keep the wavepacket on the quantum brachistochrone~\cite{Lam2021}.

\subsection{Rydberg atoms}\label{sec:Rydberg}
Rydberg atoms are highly excited atoms that exhibit a strong long-range state-dependent interaction that allows one to engineer entanglement and collective effects between two or more atoms~\cite{Saffman2010}. In particular, if one atom is prepared in a highly-excited Rydberg state, the transition to this state is shifted out of resonance for the surrounding atoms. This effect is called the Rydberg blockade and can be observed both in atomic ensembles~\cite{Heidemann2007} and in pairs of atoms~\cite{Urban2009,Gaetan2009}. In the latter case, if each of the two atoms encodes a qubit in its ground states the Rydberg blockade can be exploited to generate entangling gates~\cite{Jaksch2000,Isenhower2010,Wilk2010,Mueller2011,Mueller2014}. Ref.~\cite{Mueller2014} studied the gate scheme of Ref.~\cite{Jaksch2000}, where both atoms are excited with the same two lasers and the Rydberg blockade introduces the entangling phase, with the values from the experimental setup of Ref.~\cite{Gaetan2009,Wilk2010}. An optimization with a Gaussian pulse parametrization was used to investigate the achievable gate fidelity, approximately~$99$\,\% for a realistic scenario, as well as the required experimental improvements to achieve a gate fidelity beyond $99.9$\,\%.

Atomic ensembles exhibiting the Rydberg blockade can be used to generate single or correlated photons. In this case, the Rydberg blockade is used to generate a collective single excitation of the atoms and then collective spontaneous decay of this excitation generates the desired single photon~\cite{Dudin2012,Mueller2013,Ripka2018}. In this context, CRAB was used to optimize the state
preparation to suppress double excitations (which would lead to a second photon) and to maximize the directionality of the single photon emission~\cite{Mueller2013}. A similar process can be employed to absorb and re-emit single photons that act as flying qubits. In this way, the atomic ensemble can serve as a quantum memory~\cite{Lin2016}. Together with the previously mentioned universal quantum gates, this single photon source is a key step towards quantum computing with Rydberg atoms~\cite{Henriet2020}. Along this line, also a Rydberg quantum bus was investigated that could be used to move qubits in an optical lattice without moving the atoms~\cite{Weimer2012}. This scheme was optimized with dCRAB and it could be shown that high fidelities can be achieved for the direct transport of a qubit over up to six sites or iterative transport over larger distances~\cite{Mueller2016}.

Rydberg atoms on optical lattices can be used to study numerous interesting quantum many-body phenomena, such as quantum crystals of excited and blockaded atoms~\cite{Schauss2012} or the simulation of quantum Ising models~\cite{Labuhn2015,Schauss2018}. In Ref.~\cite{Cui2017}, it has been shown for Rydberg atoms on an optical lattice how optimal control with CRAB can be utilized to generate specific many-body states, e.g. GHZ states or symmetric superpositions of Fock states, with high fidelity. Ref.~\cite{Omran2019} designed CRAB pulses to generate a Schrödinger cat state of Rydberg atoms and demonstrated it experimentally for up to 20 qubits, the largest to date. These are essential building blocks for quantum simulators that might soon cross the threshold between pure proof of principle experiments and quantum simulations that cannot be performed classically~\cite{Scholl2020,Bloch2020}.

\subsection{Bose-Einstein condensates}\label{sec:BEC}

Since their first observation 25 years ago, Bose-Einstein condensates (BEC) have been an attractive physical platform to investigate fundamental physics, such as the measurement of atomic masses and the fine-structure constant, and to achieve many potential real-world applications, for instance in developing a portable quantum sensor for geophysics and navigation devices~\cite{Acin2018,Bongs2019}.
Various interferometric experiments have been carried out using the atom-chip technology, for example in atomic Michelson interferometry~\cite{Wang2005}, and in matter-wave interferometry in a double well~\cite{Schumm2005}. Multi-particle 
entanglement of BEC on atom chips has also been used to generate spin-squeezed states that are essential in realizing quantum metrology beyond the standard quantum limit~\cite{Riedel2010}. In scanning probe microscopes, a BEC can be used as an ultra-soft scanning probe tip, which is about a billion times softer than other AFM tips, and significantly increases the microscope's resolution~\cite{Gierling2011}. Recently, a BEC has been created in space to demonstrate atom-optics tools needed for satellite gravimetry, for quantum tests of the equivalence principle, and for gravitational-wave detection~\cite{Becker2018}. This work paves the way for scalable cold atom and photon-based quantum information devices, which can be integrated into quantum communication satellites.
Furthermore, through electromagnetically induced transparency, quantum memories 
with long storage time have been realized using BECs~\cite{Ma2017, Riedl2012, Lettner2011}.
Different quantum states and subspaces can be protected by exploiting the back action of quantum measurements, leading to quantum Zeno dynamics~\cite{Schaefer2014}.
BECs are also a promising platform for quantum simulators of strongly correlated systems to gain new insights of physics, such as high-$T_c$ superconductivity~\cite{Bloch2018}.

The CRAB algorithm was applied to BECs to realize a two-pulse atomic Ramsey interferometric measurement, which is an essential building block for atom-based quantum metrology and for investigating out-of-equilibrium dynamics at the quantum level~\cite{Frank2014}. The interferometer was realized using two optimized control pulses acting on the motional states of the condensate in an anharmonic trap, where the FOM for the optimization was the visibility of the Ramsey fringes. The pulses were obtained from open-loop CRAB optimizations performed on the dynamical simulation of the one-dimensional Gross-Pitaevski equation in the presence of intrinsic dephasing, and were able to drive the transitions between the motional states of the BEC on a timescale comparable to the trapping frequency. In the experiment, a dilute quasi one-dimensional quantum-degenerate gas of approximately 700 Rubidium-87 atoms on an atom chip at temperature $<$~50~nK was coherently controlled by the optimized CRAB pulses to create a balanced coherent superposition between the two lowest-lying motional states of the gas with a fidelity of~95\% and an interferometric contrast of~92\%. The optimization was performed using 60 Fourier components with their respective amplitudes and phases. 
The state transfer between the motional states of the BEC was later realized also at the quantum speed limit~\cite{Frank2016}. By considering finite measurement precision on the order of~1\% and taking into account experimental constraints, such as the limited bandwidth of the electronics and the maximum possible trap displacement, the optimal process reached a state transformation fidelity of approximately 99.3\% within 1.09~ms time. The QSL of state transfers between motional states of the BEC was also investigated to study the dimension of the effective Hilbert space of a quantum many-body system~\cite{Brouzos2015}. For this purpose, a BEC of interacting atoms in a double-well potential was investigated under three different approximations that alter the dimension taken into account by the simulation. When the pulse operation time approached the QSL, the optimal transfer in all three scenarios was a rotation  along the geodesic on the Bloch sphere given by the initial and target state (see Sec.~\ref{sec:many-body} and Fig.~\ref{fig:geodesic} for more details).

On a similar setup of a BEC on an atom chip, a protocol for the tomographic reconstruction of a quantum state was demonstrated based on time-resolved measurements of the atomic population distribution among the internal BEC states, implemented by an optimization protocol that minimized the difference between the measured data and the numerical simulation~\cite{Lovecchio2015}. This provides a simple yet highly advantageous tomographic method, which makes use of a post-processing optimization procedure and a standard absorption imaging technique. Building on this technique, optimal state preparation of arbitrary hyperfine superposition states was demonstrated~\cite{Lovecchio2016}. To this end, the control pulses were designed by open-loop CRAB optimization of the frequency-modulated radio-frequency~(rf) pulses using 7 random frequency basis functions and the state preparation was numerically predicted at ~$\approx$~3\% error in the final population distribution. The tomographic reconstruction developed in~\cite{Lovecchio2015} confirmed that the density matrix of the prepared state was in agreement with the theoretical prediction with a fidelity higher than~0.9, and demonstrated that the optimized rf pulses outperformed the unoptimized pulses even for shorter pulse duration, paving the way to overcoming the unavoidable and detrimental dephasing of the levels. 

Even the production of BECs themselves can be optimized in the chopped random basis. Ref.~\cite{Heck2018} used the production of a BEC from ultra-cold atoms as a benchmark to determine the performance of a gamified interface allowing hundreds of citizen scientists worldwide to actively participate in real-time optimizations by sending experimental parameters through an online cloud server and comparing the results to a remote closed-loop optimization with dCRAB (RedCRAB). The optimized experimental parameters were the magnetic field gradient and the intensity of two orthogonal dipole beams trapping the Rubidium-87 atoms. The mean number of atoms in the BEC was taken as the optimization FOM and obtained via image analysis with resonant light. The optimization with RedCRAB required only about 100~iteration steps to arrive at the optimal solution that outperformed the result obtained from the conventional hybrid trap method by more than~10\%, proving that RedCRAB is remarkably effective at finding closed-loop optimal solutions. At the same time it could also yield topological information on the underlying optimization landscape by showing that the constrained local search intrinsic to the CRAB expansion through the dCRAB super-iterations was able to find a path from good guess pulses (discovered by the citizen scientists through the gamified interface) to the global optimum.

The capability of (d)CRAB to optimize the necessary high fidelity quantum operations for sensing and metrology applications in the BEC experiments discussed above, both in open and closed-loop fashion, hints to its potential also for other technological applications with BECs, such as in satellite-based gravimetry, and for more fundamental research, for instance in gravitational-wave detection and quantum simulations of exotic states of matter. 

\subsection{Nitrogen-vacancy colour centres}\label{sec:NV}
Precise control of robust qubits at room temperature will significantly advance the development of quantum 
nanotechnology applications such as quantum improved sensing and metrology~\cite{Schmitt2017,Degen2017,
	Pezze2018,Barry2020}. Nitrogen-vacancy (NV) colour centres in diamond are a promising platform for the implementation of 
such accurate operations as the associated ground-state spin system can be initialized and read out optically, 
and controlled by microwave radiation at room temperature. The platform is also remarkably interesting in the 
context of room temperature quantum processors, where quantum logic elements are realizable by exploring 
long-range magnetic dipolar coupling between individually addressable single electron spins associated with 
separate colour centres~\cite{Neumann2010}. However, due to the unavoidable noise from the interaction with 
the environment, e.g. the nuclear spin bath, and the experimental errors such as those from apparatus imperfections, 
the microwave control needs non-trivial optimization to provide high fidelity and robust quantum operations
~\cite{Unden2016}. Such operations can be quantum state transformations or quantum gate synthesis. QOC 
methods relying on the gradient information to  obtain optimized pulses for state transformations and gate 
synthesis by means of open-loop optimization have been applied in experiments with NV centres to perform 
quantum simulation in a solid-state spin register~\cite{Wang2015}, high fidelity entanglement~\cite{Dolde2014}, 
and quantum error correction~\cite{Waldherr2014}. Generally, the application of~QOC in various essential 
quantum information processing tasks, such as high dynamic-range magnetometry via the phase estimation 
protocol~\cite{Waldherr2012}, and determination of magnetic field vector orientation~\cite{Weggler2020}, will be vital for quantum technologies based on NV centres.  

The CRAB algorithm was used for designing and optimizing a set of microwave control pulses that performs accurate state 
transformations beyond the standard rotating-wave approximation (RWA) regime in an experiment with NV centres~\cite{Scheuer2014} without the necessity to use the gradient information. The optimized microwave 
pulses were obtained by open-loop CRAB optimization where the relevant parameters were taken from experimental 
calibrations, such as the actual frequency band of the wave generator and the measured linewidth of the spin 
transition. In the experiment two essential quantum sensing building blocks, $\pi$- and 
$\pi/2$-rotations, were realized on a time scale faster than the RWA regime and a high fidelity of more than~$95\%$
(subject to experimental uncertainty of approximately~$1\%$) was reached, in agreement with the theoretical and 
numerical predictions. Furthermore, the robustness of the CRAB control pulses was further confirmed by applying 
such pulses in two standard NMR and ESR sensing sequences: free-induction decay and spin-echo sequences.  These results pave the way 
for advancing a wide range of applications, for instance for broadband magnetometry.

In another experiment with NV centres the dCRAB algorithm was embedded directly in the setup to perform single spin qubit calibrations in a real-time and autonomous way via closed-loop optimization for both quantum state transformation and quantum gate synthesis~\cite{Frank2017}. The autonomous calibrations were done in the presence of systematic errors of detuning and their performances were evaluated completely through the standard quantum 
state and gate tomography. For spin inversion, a calibrated quantum state transformation reached fidelity of  1 (with an experimental error of about 1\%) on a time 
scale of minutes, while for a single-qubit $\pi/2$-rotation a gate fidelity of 98\% was achieved within a few hours of optimization within an experimental error of approximately~2\%. This shows the potential for robust 
quantum technology applications and could be significantly improved in future by the application of faster and simpler 
quantum gate verification techniques such as randomized benchmarking~\cite{Knill2008}. Using the faster gate 
verification technique, the work could be extended to a larger dimensional quantum system to autonomously 
synthesize multi-qubit quantum registers as the algorithm does not need any prior knowledge of the system and its interaction with the environment. 

The (d)CRAB algorithm, including the remote version~RedCRAB, together with the most common gradient optimization methods, has been recently presented in an introductory review on the state-of-the-art and perspectives of 
QOC-enabled quantum operations using NV-centres~\cite{Rembold2020}. The review highlights the significance of 
QOC to drive diamond-based quantum systems fast into the realm of commercial technologies. Furthermore, the 
dCRAB algorithm and its remote version RedCRAB are straightforwardly adaptable for optimizing generic pulse 
sequences commonly used in sensing experiments with NV centres, where the optimization can be performed on 
the pulse sequences as a whole. The application of dCRAB for designing optimal filter functions of the sensing 
sequences in quantum noise spectroscopy has indeed been proposed and simulated in possible experimental 
implementations including NV-centres~\cite{Mueller2018}~(see also~Sec.~\ref{sec:enhanced-measurements}). 
A similar approach using direct optimization on sequences of $\pi$-pulses to maximize the sensing sensitivity was also 
demonstrated experimentally by Poggiali et al.~\cite{Poggiali2018}. 

NV-centres in diamond form a highly advance 
platform for sensing and the contribution of QOC to these tasks has been highlighted in~\cite{Rembold2020}. Furthermore, recent advances in diamond sample fabrication~\cite{Osterkamp2020}, and in control over a 
large number of spins associated with the NV centres~\cite{Abobeih2019, Bradley2019}, unlock new perspectives for 
the (d)CRAB method to rapidly assist and develop the realization of diamond-based quantum simulators and quantum 
registers. In the context of diamond-based quantum networks and quantum memories the (d)CRAB method is also 
straightforwardly applicable to other promising colour centres, such as silicon vacancy centres (SiV)~\cite{Nguyen2019}, 
which is a good platform for studying energy efficient control at milli-Kelvin temperature and has a memory time exceeding 
10~ms~\cite{Sukachev2017}.

\subsection{Trapped ions}
Trapped Ions are considered to be one of the most advanced quantum technology platforms, especially for quantum computation~\cite{Leibfried2003,Singer2010,Monz2011,Bermudez2017,Olsacher2020} and quantum simulation~\cite{Casanova2011,Lanyon2011,Hempel2018} as well as quantum thermodynamics \cite{Huber2008,Rossnagel2016}.
The electronic level structure of trapped ions allows to select stable qubit states that can be manipulated in a controlled way through optical~\cite{Schindler2013}, stimulated Raman~\cite{Poschinger2009} or microwave \cite{Ospelkaus2011} fields that can be used for quantum information processing purposes. While each ion corresponds to one qubit, entanglement between the ions can be achieved via Coulomb interaction or more commonly via common phonon modes in the trap. In this way, gates between the qubits can be realized, e.g.,  the Cirac-Zoller CNOT gate~\cite{Cirac1995,FSK2003}, or the M\o{}lmer-S\o{}rensen (geometric phase) gate \cite{Soerensen2000,Leibfried2003b} with laser driving, also together with methods of dynamical decoupling \cite{Bermudez2012,Tan2013}, or with microwave pulses \cite{Zarantonello2019}.

There are two basic approaches to scale up the number of qubits~\cite{Bermudez2017,Olsacher2020,Kaushal2020}: i) ion transport can be used to shuttle ions in the trap so that each pair of ions (or also larger groups of ions) can be brought close to each other and interact. ii) alternatively, in the string-based or hiding-based approach, a large number of ions is trapped in a one dimensional string and individual addressing of the ions allows selective interactions of each pair or subgroup of ions via a common axial trap mode. The two approaches can also be combined.
Ion transport has first been demonstrated almost 20 years ago \cite{Rowej2002} and shows remarkable fidelities \cite{Walther2012,Kaufmann2018}. The scalability of the approach has been studied in Ref.~\cite{Kaushal2020}. Further improvement can be expected by optimally shaped modulations of the trap potential~\cite{Fuerst2014}.
The string-based approach allows to entangle also several ions on the string at once through multi-qubit gates. In this way GHZ states of up to 14 qubits~\cite{Monz2011} have been realized.
The scalabilty for this approach has been studied in Ref.~\cite{Olsacher2020}. The potential and the required improvements to achieve fault-tolerant precision together with quantum error correction codes has been studied for both approaches in Ref.~\cite{Bermudez2017}.

Quantum measurements of the ions' quantum state can determine directly only the spin state of the ion. In many cases, however, one also needs to determine the motional state of the ion in the trap. This can be done only indirectly by entangling the motional state with the spin state. For this purpose, in Ref.~\cite{Mueller2015} CRAB was used to map the population of a given motional state onto one of the two spin-qubit states and the population of all other motional states onto the other spin-qubit state. This creates a mapping 
\begin{eqnarray}
	\ket{\downarrow}\otimes|n\rangle
	\stackrel{U_m}{\longleftrightarrow}
	\begin{cases}
		\ket{\uparrow}\otimes\sum_{n'} c^{(n,m)}_{n'}|n'\rangle\quad n=m\\
		%%%
		\ket{\downarrow}\otimes\sum_{n'} c^{(n,m)}_{n'}|n'\rangle\quad n\neq m
	\end{cases}\,,
\end{eqnarray}
where  $\ket{\downarrow},\,\ket{\uparrow}$ are the spin states and $|m\rangle,\,|n\rangle$ are the motional Fock states. The ion is initially in the $\ket{\downarrow}$ state and an arbitrary motional state. The mapping $U_m$ then maps the population of the state $\ket{m}$ onto $\ket{\uparrow}$ and the population of all other motional states onto $\ket{\downarrow}$. A subsequent measurement of the spin population gives the initial population of the motional state $\ket{m}$. By reversing the unitary mapping operation after the projective measurement, the procedure can also be used to prepare motional Fock states. The preparation and readout of motional Fock states can also be achieved through repetitive spin state measurements with partial selectivity to the motional state as has been proposed in Ref.~\cite{Huber2008}. In this reference, the technique was used to propose the demonstration of the quantum Jarzinsky equation with a trapped ion. As shown in Ref.~\cite{Mueller2015}, the application of optimal control pulses can generate the unitary mapping $U_m$ to speed-up and improve the process.

Trapped ions were studied as a possible platform for the preparation of squeezed states in Ref.~\cite{Pichler2016} (see also Sec.~\ref{sec:enhanced-measurements}). While in the absence of noise squeezed states can be prepared via adiabatic one-axis twisting, the presence of noise can prevent this approach already at a very moderate system size of a few spins (ions). For squeezed spin state preparation with trapped ions, a global magnetic field fluctuation was identified as the most significant noise source~\cite{Monz2011} and it was modeled as random telegraph noise~\cite{Pichler2016}. For up to 500 spins the state preparation protocols designed by CRAB could reproduce also in the noisy scenario the squeezing obtained by the adiabatic protocol in the noiseless case.

GHZ states of many qubits \cite{Monz2011} are also a resource for quantum  sensing~\cite{Degen2017,Szankowski2017}. The sensitivity to an external field is enhanced by a factor of $\sqrt{L}$ for a GHZ state of $L$ qubits as a result of the entanglement. Inconveniently, also the dephasing time is enhanced by the same factor and thus the creation of entanglement is not beneficial for quantum sensing directly. However, entangled states allow for sensing protocols that are not feasible with product states. Sza{\'{n}}kowski et al. \cite{Szankowski2016} showed how entangled states can be used to resolve not only temporal but also spatial correlations. In Ref.~\cite{Mueller2018} CRAB was employed to design optimized filter functions for a GHZ state of trapped ions that allow to fight spectral leakage of the signal. In particular, it was shown how an entangled state with single addressing is much more robust against high-frequency noise than a single qubit sensor and that also time-dependent spectra could be temporally resolved at a faster rate.

The studied examples have demonstrated the potential of QOC for trapped ions applications, although the proposals were only numerical and experimental implementation might need closed-loop control. In particular, trapped ions are a formidable candidate for a quantum computer platform and QOC could help to overcome the error threshold needed for quantum error correction~\cite{Bermudez2017,Olsacher2020,Kaushal2020}.

\subsection{Superconducting Qubits}
Superconducting qubits are one of the most promising platforms for quantum computing~\cite{Makhlin2001,Clarke2008,Devoret2013,Preskill2018,Wendin2017,Arute2019}. The superconducting properties introduce distinct quantum features to a macroscopic level. The quantization of charge allowed the realization of cooper pair boxes with Josephson charge qubits~\cite{Nakamura1999,Makhlin2001}, but also phase qubits and flux qubits are possible~\cite{Clarke2008}. Superconducting qubits can be embedded into a microwave resonator, leading to circuit quantum electrodynamics, and thus the qubit states can be conveniently manipulated~\cite{Wendin2017}. Also the resonator state can exhibit quantum features such as a photonic cat state~\cite{Puri2017}. A major control challenge is the potential leakage of the qubit state to levels outside the logical subspace and thus excitation pulses have to be highly frequency-selective~\cite{Motzoi2013}, especially for the transmon qubit. The transmon qubit is smilar to a Josephson charge qubit but operates in a different regime where it is less affected by charge noise at the cost of lower anharmonicity~\cite{Koch2007}. The regime of parameters can be optimized through system design~\cite{Goerz2017}.
Recent breakthroughs with superconducting qubits were the demonstration of a Schrödinger cat state with 20 artificial atoms~\cite{Song2019} and the demonstration of quantum 'supremacy', i.e., solving a problem on a quantum computer that could not be solved on a classical computer at a human time scale~\cite{Arute2019}. Ref.~\cite{Krantz2019} provides a hands-on introduction to the quantum engineering of superconducting qubits as a guide 
to the central concepts and challenges including qubit design, noise sources, and readout techniques. 

Ref.~\cite{Caneva2011} considered a pair of Josephson charge qubits modeled as perfect two-level systems with a $\sigma_z\sigma_z$-type Coulomb interaction to investigate how randomization of the basis functions and the number of basis functions for CRAB influence the convergence and the final fidelity of state preparation in this system. For a set of predefined pure and entangled states and control of the time-dependent interaction it was shown that with randomization of the basis functions already 4 CRAB parameters were sufficient to achieve errors below $1\,\%$ and for some cases machine precision was reached for only a few more parameters. Ref.~\cite{Goerz2015} also studied a pair of Josephson charge qubits, this time coupled by a Josephson junction~\cite{Siewert2001,Montangero2007} with a focus on the achievable transformations based on local equivalence classes of two-qubit gates (see Sec.~\ref{sec:quantum_gates}). The system allows for two control knobs: the Josephson coupling of the single qubits and the Josephson coupling of the junction . Two scenarios were studied: a) a single control pulse for both knobs and b) two independent control pulses for both the local driving and the interaction term. Without leakage (i.e., by considering only two levels for each qubit) the system is controllable and it could be shown that all local equivalence classes can be reached~\cite{Watts2015,Goerz2015}. For realistic parameters and considering also leakage, instead, only certain equivalence classes among the perfect entanglers could be reached with very high fidelity while other equivalence classes were not reachable. Both the maximum fidelity as well as the set of reachable gates increased by using two independent pulses. Furthermore, a transmon gate was studied in Ref.~\cite{Goerz2015}. This laid the foundation to investigate systematically the accessible universal quantum gates in transmon qubits with different parameters and to optimize the design of such parameters~\cite{Goerz2017}. An open question for superconducting qubits is the modeling of the system to an accuracy that can match the very high fidelities that can in principle be reached. To this end for a given mathematical description the system parameters have to be determined from experimental feedback. Ref.~\cite{Wittler2020} proposed a three step approach called $\mathbb{C} ^3$ (control, calibration, characterization) which combines the design of open-loop control pulses, the subsequent calibration of these pulses on the quantum device and finally an exploitation of these calibration measurements to determine the system parameters.

\subsection{Conclusions}
In this section we have given a short introduction to some of the most common quantum technology platforms with a focus on the achievements and prospects of QOC in the chopped random basis. In particular, we have presented concrete applications of the toolbox of Sec.~\ref{sec:applications-toolbox} and shown that many of the tools were already used for different platforms. Indeed, most of the tools are not only platform-independent in principle, but also in practice from a QOC point of view the engineering of quantum technology poses very similar problems for different platforms. We thus hope that this summary will contribute to a fruitful exchange of knowledge between the experts for the different platforms.

%**************************************************************
\section{Summary and Outlook}\label{sec:conclusions}
In this review, we have presented the developments of the (d)CRAB algorithm and its analytic foundations, numerical advancements, and experimental realizations from its initial conception ten years ago to the current achievements in theoretical and experimental quantum physics, engineering, and technologies.
Due to its versatility, the (d)CRAB algorithm has been applied together with several software toolboxes
for numerical simulation or experimental steering of quantum systems.
To make QOC more accessible and to lower the hurdle to future applications the algorithm is available as a software package (RedCRAB, available upon request from the authors).

To further improve the dCRAB approach to QOC toward a ready-to-use tool with high efficiency a few open questions remain.
While the random truncated basis expansion is independent of the underlying maximization or direct search algorithm, it can still have a high impact on the convergence properties of the optimization. The best choice of this underlying maximization algorithm is most likely problem-dependent and many examples of such algorithms have been given in Sec.~\ref{sec:algorithm}. Tools from machine learning could be used to systematically utilize the acquired experience for a given new QOC problem. The same arguments can be made for the choice of the basis functions.

In Sec.~\ref{sec:applications-toolbox} we have reviewed a rich toolbox of control objectives that allows to tackle many different QOC problems. However, if a full process tomography has to be avoided (as on some quantum technology platforms it is very time consuming) not all control objectives can be readily implemented in a closed-loop optimization. In these cases randomized benchmarking or similar ideas can be very important to speed up the single function evaluations of the control objectives~\cite{Wallman2016,Chasseur2017} and thus make optimization experimentally feasible. Also Bayesian methods or model learning and hybrid open- and closed-loop optimization (where the closed-loop optimization is also used to calibrate parameters of a simulation for open-loop optimization) can provide an important contribution to the overall efficacy~\cite{Wittler2020}. This argument is also related to the theory presented in Sec.~\ref{sec:constrained-optimization} that shows that the more information one is able to exploit in the design of the optimization (e.g., the knowledge of the system model or a good guess pulse), the less information has to be contained in the actual optimized parameters and thus potentially the fewer iterations of the optimization are required~\cite{Lloyd2014,Mueller2020,Gherardini2020}. Apart from various open-loop QOC techniques \cite{Jurdjevic1996,DAlessandro2007,Brif2010,Glaser2015,Koch2016}, also semi-analytic approaches like shortcuts to adiabaticity \cite{Guery-Odelin2019} can serve for this purpose.

The only way to optimize the desired quantum operation via closed-loop optimization is through measurements and processing of the measurement results (e.g., to update the control pulse for the next iteration). Potentially, these measurements can be taken also during the pulse operation time (or time evolution of the system). In this case we do not refer to it as closed-loop optimization but as feedback optimization. Weak quantum measurements can be used to update the pulse during the time evolution~\cite{Kallush2014}. Such real-time feedback is experimentally challenging and probably not feasible for all quantum technology platforms (due to the different typical time scales) but could be an interesting tool to fight slow noise whenever possible.

While dCRAB was developed as a tool for optimal control of quantum systems, it has also applications that go beyond quantum physics. Ref.~\cite{Angaroni2019} applied the algorithm to improve medical therapy: a combination of simulating cell population dynamics and employing optimal control allowed to come up with personalized therapeutic strategies in cancer patients. More specifically, a control objective was introduced allowing to personalize the drug concentration in order to optimize the efficacy of the cure while reducing the costs, especially in terms of toxicity and adverse effects. 

We hope that this review can serve the scientific community and practitioners as a map
of this versatile tool and its wide range of applications – both in discovering potentially
new insights into physics and paving the way for novel applications in quantum
technologies. Quantum information and quantum technologies have seen a tremendous
boost in the past decade that will probably even accelerate in the coming decade as
witnessed also by the recent establishment of the EU Quantum Flagship, the U.S.
National Quantum Initiative and similar programs in other countries. We firmly
think that CRAB optimization, and QOC in general, will play a crucial role in this future development
because high performance and high fidelity quantum information and technology
operations heavily depend on the optimization of control over the system, including
closed-loop. Furthermore, for closed-loop optimization of experiments the truncated
random basis is the most natural choice for optimization under realistic experimental
constraints. Indeed, the dCRAB approach allows to access the often advantageous landscape properties of QOC problems in a blackbox and gradient-free way. Here, we have provided an overview of different physical
platforms as well as optimization tasks -- such as state preparation, entanglement generation or enhanced quantum measurements -- in the chopped random basis. Practitioners in
the different fields of quantum technology – such as quantum sensing with nitrogen-vacancy centres or quantum computing with ion traps – might use it to study and
apply the control approaches that have been successfully implemented in the respective other
fields, which will stimulate the exchange among the various fields of research and emerging branches of quantum technology and engineering.

%**************************************************************
\ack
M.M.M and T.C. acknowledge funding from the European Union’s Horizon 2020 research and innovation programme under Grant Agreements No. 817482 (PASQuanS) and No. 820394 (ASTERIQS), the German Federal Ministry of Education and Research (BMBF Project No. 13N14872, VERTICONS), as well as from the Deutsche Forschungsgemeinschaft (DFG, German Research Foundation) under Germany’s Excellence Strategy – Cluster of Excellence Matter and Light for Quantum Computing (ML4Q) EXC 2004/1 – 390534769.
S.M. acknowledges support from the European Union’s Horizon 2020 research and innovation programme under the grant agreement No. 817482 (PASQuanS), by the BMBF via the QRydDemo project, and by the Italian PRIN 2017 and the CARIPARO project QUASAR.
R.S.S. and F.J. acknowledge support from the DFG (CRC1279), the European Union Project  HYPERDIAMOND, Volkswagen Foundation (Grant No. 93432), the German Federal Ministry of Education and Research (BMBF Project No. 13N14438, 13GW0281C, 13N14808, 16KIS0832, 13N14810, 13N14990), and European Research Council Synergy Grant HyperQ (Grant No. 319130).
Part of this work has been carried out during the workshop  “Quantum Simulation - from Theory to Applications” at the Erwin Schrödinger International Institute (ESI) in Vienna.

\begin{table}[t]
	\caption{Symbols}
	\begin{tabular}{ll}\br
		$J$ & control objective (landscape)\\
		$F$ & fidelity\\
		$\varepsilon$ & control error \\
		$U_f$ & unitary time evolution associated with control pulse $f(t)$\\
		$\Gamma_f$ & general time evolution map associated with control pulse $f(t)$\\
		$|\psi(t)\rangle$ & time-evolved initial state\\
		$|\xi\rangle$& initial state\\
		$|\zeta\rangle$ & target state\\
		$H(t)$ & Hamiltonian for time evolution\\
		$H_0$, $H_1$ & drift and control Hamiltonians\\
		$N_c$ & number of CRAB coefficients in one super-iteration\\
		$c_i$ & CRAB coefficients\\
		$f(t)$ & control pulse\\
		$g(t)$ & guess pulse\\
		$1/\Lambda(t)$ & shape function for pulse update\\
		$c^j_i$ & CRAB coefficients for dCRAB\\
		$f^j(t)$ & control pulse at super-iteration $j$\\
		$f^{(j)}_i(t)$ & basis functions for (d)CRAB expansion\\
		$\omega_{\mathrm{max}}$ & maximal frequency \\
		$\Delta \Omega$ & bandwidth \\
		$\mathcal{C}$ & control space\\
		$k(t)$ & pulse update\\
		$\mathcal{F}$ & filter function\\
		$N$ & dimension of system (over complex numbers)\\
		$L$ & number of constituents in many-body system\\
		$d$ & dimension of local space\\
		$D_s$ & dimension of system space (over real numbers)\\
		$D_r$ & reachable states\\
		$D_f$ & degrees of freedom in control pulse\\
		$C$ & channel capacity\\
		$P_f$ & power of signal (control pulse)\\
		$\hat{f}$ & power spectral density of signal\\
		$P_n$ & power of noise\\
		$\hat{n}$ & power spectral density of noise\\
		$I_f$ &  information contained in the control pulse \\
		$f_{\mathrm{max}}$ & maximal value/variation of the control pulse \\
		$\Delta f$ & resolution of the control pulse\\
		$\Delta$  & spectral gap\\
		$T_{\mathrm{QSL}}$ & quantum speed limit\\
		$U$ & unitary time evolution\\
		$V$& target unitary (gate)\\
		$A$ & non-local content of two-qubit gate\\
		$\sigma_x$, $\sigma_y$, $\sigma_z$ & Pauli matrices/operators\\
		\br
	\end{tabular} 
\end{table}

\section*{References}
\bibliographystyle{iopart-num}
\bibliography{CRABRefs}

\end{document}